\documentclass[preprint,showpacs,preprintnumbers,amsmath,amssymb,superscriptaddress, nofootinbib]{revtex4}  
\usepackage{graphicx,color}
\usepackage{amsmath,amssymb}
\usepackage{url}
\usepackage{epstopdf}
\usepackage{slashed}

\newcommand{\be}{\begin{equation}}
\newcommand{\ee}{\end{equation}}
\newcommand{\bea}{\begin{eqnarray}}
\newcommand{\eea}{\end{eqnarray}}

\newcommand{\crn}{\nonumber \\}

\newcommand{\fr}{\frac}

\newcommand{\bc}{\begin{center}}
	\newcommand{\ec}{\end{center}}

\newcommand {\ba}{\begin{array}}
	\newcommand {\ea}{\end{array}}
\newcommand{\ben}{\begin{enumerate}}
	\newcommand{\een}{\end{enumerate}}
	
\usepackage{epsfig,graphicx} 
\usepackage{bm}
\usepackage{dcolumn}
\begin{document}

\title{Charged lepton flavor violating decays  and $(g-2)_{e_a}$ in an extended standard model  with singlet and triplet leptoquarks}

\author{L.T. Hue}\email{lethohue@vlu.edu.vn}
\affiliation{Subatomic Physics Research Group, Science and Technology Advanced Institute, Van Lang University, Ho Chi Minh City, Vietnam}
\affiliation{Faculty of Applied Technology, Van Lang School of Technology, Van Lang University, Ho Chi Minh City, Vietnam}
\author{N.V. Hop}\email{nvhop@ctu.edu.vn}
\affiliation{Department of Physics, Can Tho University,
	3/2 Street, Can Tho, Vietnam}
\author{Vo Quoc Phong}\email{vqphong@hcmus.edu.vn}
\affiliation{Department of Theoretical Physics, Faculty of Physics and Engineering Physics, University of Science, Ho Chi Minh City, Vietnam}
\affiliation{Vietnam National University, Ho Chi Minh City, Vietnam}
\author{N.H.T. Nha \footnote{corresponding author}}\email{nguyenhuathanhnha@vlu.edu.vn}
\affiliation{Department of Theoretical Physics, Faculty of Physics and Engineering Physics, University of Science, Ho Chi Minh City, Vietnam}
\affiliation{Vietnam National University, Ho Chi Minh City, Vietnam}
\affiliation{Subatomic Physics Research Group, Science and Technology Advanced Institute, Van Lang University, Ho Chi Minh City, Vietnam}
\affiliation{Faculty of Applied Technology, Van Lang School of Technology, Van Lang University, Ho Chi Minh City, Vietnam}

\begin{abstract}
We study the anomalous magnetic moments of charged leptons and their lepton flavor-violating decays in a Standard Model extension containing one $SU(2)_L$ singlet and one $SU(2)_L$ triplet scalar leptoquark. Our results reveal significant correlations among $\Delta a_{e,\mu}$ and the decay rates of $e_b\to e_a\gamma$, and $h,Z\to e_b e_a$. In particular, the branching ratios $\mathrm{Br}(e_b\to e_a\gamma)$ and $\mathrm{Br}(h,Z\to e_b e_a)$ exhibit significant correlations. The model cannot simultaneously accommodate sizable values of $|\Delta a_e|=\mathcal{O}(10^{-13})$ and $|\Delta a_\mu|=\mathcal{O}(10^{-10})$. For $|\Delta a_\mu|=\mathcal{O}(10^{-10})$ and $\mathrm{Br}(\mu \to e\gamma)>10^{-15}$, the decay rates  are suppressed to $\mathrm{Br}(\tau \to e\gamma)<\mathcal{O}(10^{-12})$ and $\mathrm{Br}(h\to \tau e)<\mathcal{O}(10^{-6})$, while the remaining ones  can reach the forthcoming experimental sensitivities. Conversely, for $|\Delta a_e|=\mathcal{O}(10^{-13})$, one obtains the suppressed decay rates  $\mathrm{Br}(\tau \to \mu \gamma)<\mathcal{O}(10^{-11})$ and $\mathrm{Br}(h\to \tau \mu)<\mathcal{O}(10^{-7})$.

%\\\textbf{Last updated: 09, Sep, 2026.........., delete this before submitting this manuscript to journal}
\end{abstract}
%\pacs{ %11.15.Ex  Supersymmetric models}

%\subjectindex{B40, B50, B53, B54, B56}

\maketitle
%%%%%%%%%%%%%%%%%%%
\section{\label{intro} Introduction}
\allowdisplaybreaks
Scalar leptoquarks (LQs) provide a well-motivated framework for connecting the quark and lepton sectors and have been extensively studied in anomalous magnetic moments (AMMs) $ a_{e_a} \equiv (g-2)_{e_a}/2$ of charged leptons $e_a$ ($e_a$AMMs) and charged-lepton flavor-violating (cLFV) processes, with $\mu\to e\gamma$ being one of the most extensively studied channels \cite{Anselm:1985bp,Cheung:2001ip,Mahanta:2001yc,Andreev:2006wh,Benbrik:2008si,Benbrik:2010cf,Crivellin:2017zlb,Buttazzo:2017ixm,Marzocca:2018wcf,Arnan:2019olv,Crivellin:2020mjs,Dorsner:2020aaz,Crivellin:2021ejk,FileviezPerez:2021lkq,Parashar:2022wrd,Freitas:2022gqs,Crivellin:2022mff,Khasianevich:2023duu,Dev:2024tto,De:2024foq,Nha:2026yat}. Among these, models containing both a singlet and a triplet scalar LQ have been studied in various phenomenological contexts \cite{Crivellin:2019dwb,Gherardi:2020qhc,Bordone:2020lnb,DaRold:2020bib,Marzocca:2021miv}. More recently, the singlet--triplet scalar LQ framework has been studied in Refs.~\cite{Saad:2020ihm,Greljo:2021xmg,Chen:2022hle,Bhaskar:2022vgk} in connection with the muon $(g-2)$ and the $R_{K^{(*)}}$ and $R_{D^{(*)}}$ anomalies, with different studies addressing different combinations of these observables. In particular, the chirality-flipping contribution involving the top quark plays an important role in accommodating the muon $(g-2)$ anomaly \cite{Bhaskar:2022vgk}. However, the analysis have focused primarily on these anomalies and have not investigated $e_a$AMMs together with LFV processes, including cLFV $e_b\to e_a\gamma$, SM-like Higgs (LFV$h$) $h\to e_be_a$, and $Z$ boson (LFV$Z$) $Z\to e_b^\pm e_a^\mp$.

Furthermore, the experimental situation makes such an analysis timely. The stringent upper limits on cLFV decays, especially $\mu\rightarrow e\gamma$, place strong restrictions on flavor-changing couplings, while searches for LFV$h$ and LFV$Z$ decays provide relevant constraints. In particular, the current experiments have pointed out the following upper bounds of the LFV decay rates:
\begin{align}\label{LFV_exp}
&\bullet\; \text{cLFV decay rates \cite{BaBar:2009hkt, MEG:2016leq, Belle:2021ysv, MEGII:2023ltw, MEGII:2025gzr}:}\; \mathrm{Br}(\mu\rightarrow e\gamma) < 1.5\times 10^{-13}, \; \mathrm{Br}(\tau\rightarrow \mu\gamma) <4.2\times 10^{-8},\crn
&\mathrm{Br}(\tau\rightarrow e\gamma) <3.3\times 10^{-8};
\crn
&\bullet\; \text{LFV\textit{h} decay rates \cite{CMS:2021rsq, ATLAS:2019xlq, CMS:2023pte, ATLAS:2023mvd}:}\; \mathrm{Br}(h\rightarrow \mu e)<4.4\times 10^{-5},\; \mathrm{Br}(h\rightarrow \tau\mu)<1.5\times 10^{-3},\crn
& \mathrm{Br}(h\rightarrow \tau e)<2.0\times 10^{-3};
\crn
&\bullet\; \text{LFV\textit{Z} decay rates \cite{ATLAS:2021bdj, ATLAS:2022uhq, CMS:2025wqy}:}\; \mathrm{Br}(Z\to \mu^\pm e^\mp) \leq 1.9\times 10^{-7},\; \mathrm{Br}(Z\to \tau^\pm e^\mp) \leq 5.0\times 10^{-6},\crn
& \mathrm{Br}(Z\to \tau^\pm\mu^\mp) \leq 6.5\times 10^{-6}.
\end{align}

At the same time, the recent measurement of the  $\mu$AMMs with improved precision further motivates the study of possible new-physics contributions to $(g-2)_\mu$ \cite{Muong-2:2023cdq,Muong-2:2025xyk, Muong-2:2006rrc}. The discrepancy between the SM prediction based on lattice-QCD \cite{Borsanyi:2020mff, RBC:2024fic, Djukanovic:2024cmq, Boccaletti:2024guq} and the new experimental value was found \cite{Aliberti:2025beg} to be $\Delta a_\mu \equiv a_\mu^{\mathrm{SM}}-a_\mu^{\mathrm{exp}} = (3.8\pm6.3)\times10^{-10}$. The electron AMM provides another sensitive constraint on new physic contributions, owing to the high precision with which the electron AMM and the
fine-structure constant have been measured \cite{Hanneke:2008tm,Parker:2018vye,Morel:2020dww,Fan:2022eto}.
These experimental inputs motivate a simultaneous investigation of LFV decays and $e_a$AMMs, since the same LQ-fermion couplings can contribute to both classes of observables at the one-loop level.

These considerations motivate a closer examination of the interplay between the two LQ representations. Although the two LQs have different $SU(2)_L$ quantum numbers and therefore distinct component fields and fermionic couplings, their simultaneous presence provides additional structures in the one-loop amplitudes.  
This interplay and mixing may lead to interesting correlations among these observables that are absent in a single LQ framework. We therefore investigate the combined framework of the model with two LQs (called the LQST model for short) studied in Ref. \cite{Bhaskar:2022vgk} through radiative cLFV, LFV$h$, LFV$Z$ decays, and the $e_a$AMMs, and examine whether their combined contributions can satisfy the current experimental constraints and account for the observed deviations in $e_a$AMMs.

The paper is organized as follows. In Sec.~\ref{sec_model}, we review the structure of the LQST model and introduce the particle spectra in this model. In Sec.~\ref{sec_coupling}, we present the relevant interactions and derive the general analytical expressions for the one-loop contributions to LFV processes and anomalous magnetic moments. In Sec.~\ref{sec_numerical}, we perform a comprehensive numerical analysis and discuss the resulting phenomenological implications. Finally, our conclusions and several highlights obtained from this framework are given in Sec.~\ref{conclusion}. Furthermore, in Appendix.~\ref{app:higgs}, we show the general Higgs potential used in this model. Besides, we will summarize all of the detailed calculations relevant to determining the masses and the mixing matrix of the LQs.

\section{\label{sec_model} The SM model with  singlet and triplet leptoquarks}

By adding the two new LQs, including the singlet $S$ and triplet $T=\left(T^{q_1}, T^{q_2}, T^{q_3}\right)^\mathrm{T}$ into the SM model \cite{Bhaskar:2022vgk}, they provide a direct connection between the quark and lepton sectors \cite{Buchmuller:1986zs,Crivellin:2020mjs,He:2026nul} in the LQST model.  Unlike the Higgs scalar fields, which are singlets $SU(3)_C$, the LQs are $SU(3)_C$ antitriplets, leading to they have color interaction. The specific expression representing the multiplets' components according to the standard group $SU(3)_C\otimes SU(2)_L\otimes U(1)_Y\otimes Z_2$ is shown in Fig. \ref{eq:331LQhat}.
\begin{table}[ht]
\centering
\begin{tabular}{|c|c|c|c|c|c|c|c|c|c|c|}
\hline
& $ L_{aL}$ & $e_{aR}$ & $Q_{1,2L}$ & $Q_{3L}$  & $u_{R}, c_R$ & $t_{R}$  & $d_{aR}$ &  $\phi$ & $S$ & $T$ \\
\hline
$SU(3)_C$ & 1 & 1 & 3 & 3 &  3 & 3 & 3 &  1 & $3^*$ & $3^*$ \\
\hline
$SU(2)_L$ & 2 & 1 & 2 & 2 &   1 & 1 & 1 &  2 & 1 & 3 \\
\hline
$U(1)_Y$ & $-\frac{1}{2}$ & $-1$ & $\frac{1}{6}$ & $\frac{1}{6}$   & $\frac{2}{3}$ & $\frac{2}{3}$ & $-\frac{1}{3}$ &  $\frac{1}{2}$ & $\frac{1}{3}$ & $\frac{1}{3}$ \\
\hline
$Z_2$ & + & + & + & $-$  & + & $-$ & $-$ &  + & $-$ & $-$ \\
\hline
\end{tabular}
\caption{Table of particle contents in the LSQT model, where $L_{aL}=(\nu_a,\; e_a)_L^T$, and $Q_{aL}=(u_a,\; d_a)^T_L$ with  $a=1,2,3$ are the fermion generation indices defined in the SM.}\label{eq:331LQhat}
\end{table}
where, $a=1,2,3$ are the fermion generation indices defined in the SM. The charge operator is defined as $\hat{Q}=\hat{T}_3+ \hat{Y}$, and the triplet LQ $SU(2)_L$ is $\hat{T}=(T^1, \;T^2,\;T^3)^T$ is transformed into a matrix representation of $2\times 2$ according to the relationship $T=\sum_{a=1}^3 \frac{\sigma_a}{2}T^a$ defined in Table.~\ref{eq:331LQhat}, with Pauli matrices $\sigma_a$. The LQ triplet components can be written using the following charge notation:
\begin{equation}
	\label{eq:S3}
T= \begin{pmatrix}	\frac{T^{\frac{1}{3}}}{\sqrt{2}}& T^{\frac{4}{3}} \\
	T^{-\frac{2}{3}}&-\frac{T^{\frac{1}{3}}}{\sqrt{2}}
\end{pmatrix}.
\end{equation}
As discussed in Ref.~\cite{Bhaskar:2022vgk}, chirality-flipping contributions involving heavy quarks, in particular the top quark, can give sizable effects in the $e_a$AMMs due to the large mass of this quark. In our analysis, we retain the top-quark couplings to account for such potentially enhanced contributions, while the down-quark couplings are also retained because the down-type quark participates in the interactions with the LQ triplet  $T$. To avoid introducing a large number of unnecessary flavor couplings and to simplify the numerical analysis, we impose an additional discrete $Z_2$ symmetry to restrict the allowed LQ--quark interactions, see Table.~\ref{eq:331LQhat}.

The Yukawa interaction Lagrangian extending the SM and satisfying the renormalizability and symmetry invariance requirements under  $SU(3)_C\otimes SU(2)_L\otimes U(1)_Y\otimes Z_2$ is given by:
\begin{align}
	\label{eq:LsqL}
	\mathcal{L}_Y^{S}=& \sum_{a,b=1}^3 \left[-x^L_{ab} \overline{(Q_{aL})^C}(i\sigma_2)L_{b}  S -x^R_{ab} \overline{(u_{aR})^C}Se_{bR} + y^L_{ab}\overline{(Q_{aL})^C} \left( i\sigma_2\right)TL_{b} \right] +\mathrm{h.c.},
\end{align}
where $(Q_{aL})^C=C \overline{Q_{aL}}^T$ is the charge conjugate of the doublet $Q_{aL}$, and $C$ is the charge conjugate operator. Since the triplet LQ component $T^{-2/3}$ couples only to  up-type quarks and neutrinos, it does not contribute to one-loop amplitudes with external charged leptons. Therefore, only the triplet LQ components with electric charges $Q(T^{1/3})=1/3$ and $Q(T^{4/3})=4/3$ are relevant for the calculations presented below.

The gauge interactions between the scalar LQs and the SM gauge bosons are governed by the covariant kinetic Lagrangian as follow:
\begin{align}
	\label{eq:lds}
	\mathcal{L}_D^S= &(D_{\mu}S)^* (D^{\mu}S)  + \mathrm{Tr}\left[(D_{\mu}T)^\dagger (D^{\mu}T) \right], \crn
D_{\mu}S=& \left(\partial_{\mu} -ig' B_{\mu}\times \frac{1}{3}\right)S,
\crn D_{\mu}T=&\partial_{\mu}T -ig \left[T, W_{\mu}\right]-i g' B_{\mu}\times \frac{1}{3}T,
\end{align}
where $g'/g = s_W/c_W\, (\text{with} \,s_W^2= \text{sin}^2{\theta_W} \approx 0.231)$, and the $SU(2)_L$ gauge field in the two-dimensional matrix representation is defined by $W_{\mu}\equiv \sum_{a=1}^{3} \frac{\sigma^{a}}{2}W_{\mu}^{a}$ with $\sigma^a$ denoting the Pauli matrices and $W_\mu^a,\, B_\mu$ is the  $U(1)$ gauge boson of the SM.

The one-loop contributions to the charged lepton AMMs as well as the cLFV, LFV$h$, and LFV$Z$ amplitudes arise from the Yukawa interactions between leptons and scalar LQs in Eq.~\eqref{eq:LsqL}. Consequently, the following discussion will concentrate on this interaction sector. To derive the corresponding one-loop amplitudes, one first needs to identify the physical scalar LQ mass eigenstates and the associated mixing parameters, which detail caclulated and presented in Appdendix.~\ref{app:higgs}.

To determine couplings of gauge bosons and SM quarks in this framework model, we expand the expression for the kinetic Lagrangian of quarks and the Yukawa Lagrangian, namely:
  \begin{align}
  \label{eq:lkquark}
  \nonumber\mathcal{L}^{qqV}=&\sum_{a=1}^3 \left[  i \overline{Q_{a L}} \gamma^\mu D_\mu Q_{a L} +  i \overline{u_{a R}} \gamma^\mu D_\mu u_{a R} +i \overline{d_{a R}} \gamma^\mu D_\mu d_{a R} \right] \crn
  =&\sum_{a=1}^3 g\left[\overline{\hat{u}}_{a}\left(\frac{3c^2_W -s^2_W}{6c_W}\gamma ^\mu P_L + \frac{2s^2_W}{3c_W}\gamma^\mu P_R\right)Z_\mu \hat{u}_{a}\right.\crn
&\left.\qquad  +\overline{\hat{d}}_{a}\left(\frac{3c^2_W +s^2_W}{6c_W}\gamma ^\mu P_L +\frac{-s^2_W}{3c_W}\gamma^\mu P_R \right)Z_\mu\hat{d}_{a}\right] +  ...,
  \end{align}
where we ignore couplings give one-loop not contributions to $(g-2)_{e_a}$ and LFV decays. In addition, the physical basis of quark $\hat{q}_a$  with $q=u,d$ relate to the flavor basis  through the following relation:   $\vec{q}_{L(R)} = V^{q\dagger}_{L(R)} \vec{\hat{q}}_{L(R)}$, where $\vec{q}=(q_1,q_2,q_3)^T$ and $V^{q\dagger}_{L(R)}$ are $3\times3$ unitary matrices used to diagonalize the two up- and down-quark mass matrices. In the following calculations, we adopt the usual assumption that $V^{u}_R=V^{d}_{L(R)}=I_3$ and $V^{u}_L=V_{\mathrm{CKM}}$, consistent with Ref. \cite{Bhaskar:2022vgk}.

\section{\label{sec_coupling} Couplings and analytic formulas for $(g-2)_{e_a}$ anomalies and LFV decay rates}
\subsection{\label{AMM&cLFV} $(g-2)_{e_a}$ anomalies and $e_b\to e_a\gamma$ decays}
Starting from the Yukawa interaction Lagrangian in Eq.~\eqref{eq:LsqL}, in terms of the physical states of the LQs given in Eq.~\eqref{eq:STphy}, together with the transformations into physical up-quark states 
$\overline{(u_L)^C} = \overline{(\hat{u}_L)^C}V_{\mathrm{CKM}}^T$,
we obtain
\begin{align}
 	\label{eq:LYSp}
 	\mathcal{L}_Y^{S}=&  - \overline{\vec{\hat{u}}^C} \left[ V^{T}_{\mathrm{CKM}}\left( x^Lc_{\theta}+  s_{\theta}\frac{y^L}{\sqrt{2}} \right) P_L  +  c_{\theta} x^RP_R  \right] \vec{e}\;S_-
 \crn & - \overline{\vec{\hat{u}}^C} \left[ V^{T}_{\mathrm{CKM}}\left( - s_{\theta}x^L +   c_{\theta} \frac{y^L}{\sqrt{2}}\right) P_L +s_{\theta} x^RP_R  \right] \vec{e}\;S_+ 
 	%
%\crn &
 -\overline{\vec{\hat{d}}^C} y^LP_L\vec{e} \; T^{4/3} 
 	+\mathrm{h.c.}+\dots,
 \end{align}
where $\vec{f}=(f_1,f_2,f_3)$.
Eq.~\eqref{eq:LYSp} contains the interaction vertices relevant to the one-loop contributions to the  $e_a$AMMs and cLFV decay amplitudes in the LQST model. 

 As presented in Sec.~\ref{sec_model}, in the LQST model under consideration, only the interaction terms in Eq.~\eqref{eq:LYSp} contribute to the $e_a$AMMs at the one-loop level.  The relevant one-loop diagrams involve the LQ states $S_\pm$ with electric charge  $Q_{S_\pm}=\pm1/3$ and the LQ state $T^{4/3}$ with electric charge $\pm4/3$.  The general conventions for interaction Lagrangians relevant to the one-loop contributions of scalar and vector fields to the $e_a$AMMs are given in Ref.~\cite{Crivellin:2018qmi} and are also applicable to cLFV amplitudes. Accordingly, the interaction coefficients appearing in Eq.~\eqref{eq:LYSp} are grouped into three classes corresponding to the LQ states:
\begin{align}
\label{eq:Gaxy}
&\Phi=S_-, \;F= \hat{u}^C_i,\; Q_F=- \frac{2}{3}:
\Gamma_{ia S_-}^{ L} = \left[V^{T}_{\mathrm{CKM}}\left( c_{\theta}x^L+  s_{\theta}\frac{y^L}{\sqrt{2}} \right) \right]_{ia},
\;  \Gamma_{ia S_-}^{ R} = c_{\theta}x^R_{ia}, 
\crn &\Phi=S_+, \;F= \hat{u}^C_i,\; Q_F=-\frac{2}{3}:
   \Gamma_{ia S_+}^{ L} =\left[ V^{T}_{\mathrm{CKM}}\left( - s_{\theta}x^L +   c_{\theta} \frac{y^L}{\sqrt{2}}\right)\right]_{ia} , \; \Gamma_{ia S_+}^{ R} = s_{\theta}x^R_{ia}, 
\crn& \Phi=T^{4/3}, \;F= \hat{d}^C_i,\; Q_F=\frac{1}{3}:  \Gamma_{ia T}^{ L} =  y^L_{ia}, \;    \Gamma_{iaT}^{ R} =0,
\end{align}
where  $\hat{u}^C_i=u^C,c^C,t^C$ and  $\hat{d}^C_i=d^C,s^C,b^C$, with $i=1,2,3$ being the family index.

\begin{figure}[ht]
	\centering 
	\includegraphics[trim=1cm 14.5cm 1cm 3cm, clip, width=1\textwidth]{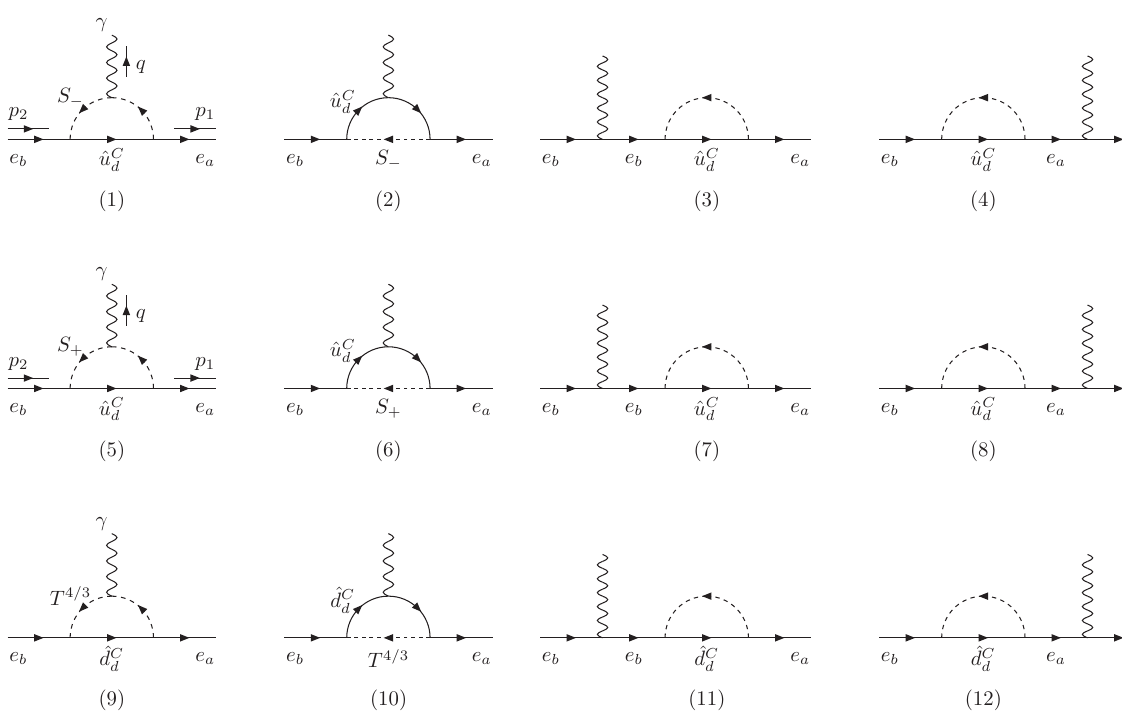}
	\caption{ One-loop Feynman diagrams with LQ exchanges contributing to $(g-2)_{e_a}$, and cLFV decay amplitudes predicted by the LQST framework, where $a, b, d=1,2,3$. }\label{f:1loopLQS13}
\end{figure}

From there, we proceeded to draw the one-loop Feynman diagram corresponding to these contributions is presented in Fig.~\ref{f:1loopLQS13}.
 Correspondingly, the one-loop contributions to $a_{e,\mu}$ and the cLFV branching ratio (Br) predicted by the   LQST are given by \cite{Lavoura:2003xp, Lindner:2016bgg, Hue:2017lak, Crivellin:2018qmi, Tran:2022cwh}:
\begin{align}
	\label{eq:cLFV}
	a_{e_a} =& -\frac{4m_a}{e} \mathrm{Re} \left[c_{(aa)R}\right],  \\
	\mathrm{Br}(e_b\to e_a \gamma) =& \frac{48 \pi^2}{G_F^2m_b^2} \left(|c_{(ab)R}|^2 +|c_{(ba)R}|^2\right) \mathrm{Br}(e_b\to e_a \overline{\nu_a} \nu_b),
\end{align}
with $G_F=g^2/(4\sqrt{2}m_W^2)$ is the Fermi constant, $ \mathrm{Br}(\mu \to e\overline{\nu_e} \nu_\mu) \simeq 1, \; \mathrm{Br}(\tau \to e\overline{\nu_e} \nu_\tau) \simeq 0.1782,\; \mathrm{Br}(\tau \to \mu\overline{\nu_\mu} \nu_\tau) \simeq0.1739, $ \cite{ParticleDataGroup:2024cfk}. The one-loop factors $c_{(ab)R}$ are given by  $$c_{(ab)R}=\sum_{\Phi=S_{\pm},T^{4/3}} c_{(ab)R}(\Phi),$$ 
where $c_{(ab)R}(\Phi)$ are derived from one-loop Feynman diagrams given in Fig. \ref{f:1loopLQS13}.
 
The specific expression for the sum of the one-loop contribution coefficients is:
\begin{align}
\label{eq:cabRLQ}
c_{(ab)R}=&\frac{3 e}{16\pi^2 }\sum_{\Phi=S_{\pm}}\sum_{i=1}^3 \left\{   \frac{ g_{ab,i\Phi}^{LR} f_{S,u} \left( t_{i,\Phi}\right)}{m_{\Phi}}  +   \frac{\left(  m_{b} g_{ab,i\Phi}^{LL} + m_{a} g_{ab,i\Phi}^{RR} \right)  \tilde{f}_{S,u}\left( t_{i,\Phi}\right)}{m_{\Phi}^2} \right\}
\crn & +\frac{3 e}{16\pi^2 m_{T^{4/3}}^2}\sum_{i=1}^3m_{b}g_{ab,iT}^{LL}g_{S,d}\left( t_{i,T}\right) ,
\end{align}
where $t_{i,\Phi}\equiv m^2_{q_i}/m^2_{\Phi}$, namely  $t_{i,S_\pm} =  \frac{m^2_{u_i}}{M^2_{S_\pm}}$ and $t_{i,T}=\frac{m^2_{d_i}}{M^2_{T^{4/3}}}$. Here, we note that $\Phi =S_{\pm}$ and $\Phi=T^{4/3}$ correspond to $q_i=u_i$ and $q_i=d_i$, respectively. This convention will  be applied from now on.

 The factor $g_{ab,i\Phi}^{XY}$ with $X, Y = L, R$ are LFV sources derived from Eq.~\eqref{eq:Gaxy}  read 
\begin{align}
\label{eq:gXLR}
g_{ab,i\Phi}^{XY} \equiv \Gamma_{ai\Phi}^{X*} \Gamma_{bi\Phi}^{Y}.
\end{align}
The one-loop functions appearing in Eq. \eqref{eq:cabRLQ} for up- and down quarks  are defined as \cite{Crivellin:2018qmi}
\begin{align}
\label{eq:fSud}
f_{S ,u}(x)&=   x^{\frac{1}{2}}\left[ \frac{x^2-1-2 x \ln x}{4(x-1)^3} -  \frac{x-1-\log x}{3(x-1)^2}  \right] ,
\crn g_{S,u}(x)&= \frac{2 x^3+3 x^2-6 x+1-6 x^2 \ln x}{24(x-1)^4} -\frac{x^2-1-2 x \ln x}{12 (x-1)^3},
\crn   g_{S,d}(x)&= \frac{2 x^3+3 x^2-6 x+1-6 x^2 \ln x}{24(x-1)^4} + \frac{x^2-1-2 x \ln x}{24(x-1)^3}.
\end{align}
We note several important properties relevant to the numerical investigation of the parameter space regions that allow sizable LFV decay rates and $\Delta a_{e_a}$. First, we consider the upper bound  Br$(\mu \to e \gamma)< 1.5\times 10^{-13}$ corresponding the most stringent experimental constraint. Assuming that the chirality-enhanced contributions dominate, the simple estimate given in Ref.~\cite{Crivellin:2018qmi}, Br$(\mu \to e\gamma)\varpropto |\Delta a_{\mu} \Delta a_e| $, leads to the upper bound $ |\Delta a_{\mu} \Delta a_e|<\mathcal{O}(10^{-30})$. This bound therefore prevents both $\Delta a_{e}$ and $\Delta a_{\mu}$ from simultaneously taking sizable values consistent with the experimental data. We investigate the relation among these three quantities in the model under consideration.

Since the chirality-enhanced  contributions, which are proportional to left-right (LR) coupling products, are dominant only for loops involving heavy fermions, such as the top quark, we consider only the case $i=3$ for the LFV couplings in Eq.~\eqref{eq:Gaxy}. If $x^R_{3a}=0$, the chirality-enhanced contribution to $\Delta a_{e_a}$ vanishes, leaving only small contributions from the $g^{LL}$ terms, particularly for heavy LQ masses. We therefore focus on the case $x^R_{3a}\neq 0$ for all $a=1,2,3$, with these couplings sufficiently large to dominate $|c_{(ab)R}|$. In particular, for $\Delta a_{e_a}$ and Br$(\mu \to e\gamma)$, we have:
\begin{align}
\Delta a_{e_a} \simeq &	\Delta a^{LR}_{e_a} = -\frac{4m_a}{e}\mathrm{Re}[c_{(aa)R}] \left[g^{LL}=0,g^{RR}=0\right]
	\crn\simeq&  -\frac{3m_ax_{3a}^{R}}{4\pi^2} \left[   \frac{ \Gamma_{3aS^+}^{L*}s_{\theta}  f_{S,u} \left( t_{3,S^+}\right)}{m_{S^+}} +\frac{ \Gamma_{3aS^-}^{L*}c_{\theta}  f_{S,u} \left( t_{3,S^-}\right)}{m_{S^-}}   \right], \label{eq:aeaLR}
	\\ \mathrm{Br}(\mu \to e\gamma) \varpropto &\left(|c_{(12)R}|^2 +|c_{(21)R}|^2\right): 
	\left\{ c_{(12)R} ,\; c_{(21)R} \right\}\simeq \frac{-e}{4m_{e}}\times \left\{ \frac{x^R_{32}\Delta a_{e}}{x^R_{31}}, \;
	\frac{x^R_{31}\Delta a_{\mu}}{x^R_{32}} \right\}.
	\label{eq:emuLFV}
\end{align}
Therefore, if $x^R_{3a}\neq 0$ for all $a=1,2,3$ and the chirality-enhanced contributions dominate, the following approximate relation holds:
\begin{align}
	\label{eq:cLFVLR}
	\mathrm{Br}(e_b\to e_a \gamma)^ \mathrm{LR} =& \frac{12 \pi^3\alpha_e}{G_F^2m_b^2m_a^2}   \left(\left|\frac{x^R_{3b}\Delta a_{e_a}}{x^R_{3a}}\right|^2 +\left|\frac{x^R_{3a} m_a\Delta a_{e_b}}{x^R_{3b} m_b}\right|^2\right) \mathrm{Br}(e_b\to e_a \overline{\nu_a} \nu_b),
\end{align}
In particular with numerical values of lepton masses given in Ref. \cite{ParticleDataGroup:2026mpi}, as discussed in Sec.~\ref{sec_numerical}, we obtain
\begin{align}
\label{eq:cLFV1}
\mathrm{Br}(\mu \to e \gamma) \simeq & 7\times 10^{18}  \left(\left|\frac{x^R_{32}\Delta a_{e}}{x^R_{31}}\right|^2 +\left|\frac{x^R_{31}}{x^R_{32} }\times  0.005\Delta a_{\mu}\right|^2\right),
\crn \mathrm{Br}(\tau\to e \gamma) \simeq& 4.34\times 10^{15}  \left(\left|\frac{x^R_{33}\Delta a_{e}}{x^R_{31}}\right|^2 +\left|\frac{x^R_{31}}{x^R_{33} }\times 2.9\times 10^{-4}\Delta a_{\tau}\right|^2\right),
\crn \mathrm{Br}(\tau \to \mu \gamma) \simeq& 10^{11} \left(\left|\frac{x^R_{33}\Delta a_{\mu}}{x^R_{32}}\right|^2 +\left|\frac{x^R_{32}}{x^R_{33} } \times 0.06\Delta a_{\tau}\right|^2\right).
\end{align}

These relations can then be used to estimate upper bounds on$|\Delta a_{e,\mu}|$ by assuming Br$(\mu \to e\gamma) \simeq  6\times 10^{-14}$, corresponding to the sensitivity expected from forthcoming experiments \cite{MEGII:2018kmf, Belle-II:2018jsg}. Namely, if $|x^R_{32}/x^R_{31}|\ge1$, we obtain $|\Delta a_{e}| \le 10^{-16}$, while no corresponding upper bound on $|\Delta a_{\mu}|$ is obtained. In contrast, for  $|x^R_{32}/x^R_{31}|<1$, we obtain  $|\Delta a_{\mu}|<2\times 10^{-14}$. Consequently, the LQST model cannot simultaneously accommodate sizable values of both$|\Delta a_e|\varpropto \mathcal{O}(10^{-13})$ and $|\Delta a_{\mu}|\varpropto \mathcal{O}(10^{-10})$, as suggested by the current experimental data. The corresponding predictions for Br$(\tau \to \mu\gamma,e\gamma)$ are less restrictive but remain useful for cross-checking the numerical results.

The above qualitative estimates are consistent with previous discussions. Namely, although the relation between Br$(\mu \to e\gamma)$ and $\Delta a_{e,\mu}$ is not as simple as in minimal SM extensions with a single new particle that couples to both the electron and muon \cite{Crivellin:2018qmi,Athron:2025ets}, a similar conclusion holds in the LQST model with two LQs considered here. This is because the model contains only one Yukawa interaction term describing the couplings of the LQs to the right-handed charged leptons $e_{aR}$, as given in the Lagrangian in Eq.~\eqref{eq:LsqL}. This structure leads to the simple relation given in Eqs.~\eqref{eq:cLFVLR}.

\subsection{\label{LFVh} $h\to e_b^\pm e_a^\mp$ decays}
\begin{figure}[ht]
	\centering 
	\includegraphics[trim=1cm 18.5cm 1cm 3cm, clip, width=1\textwidth]{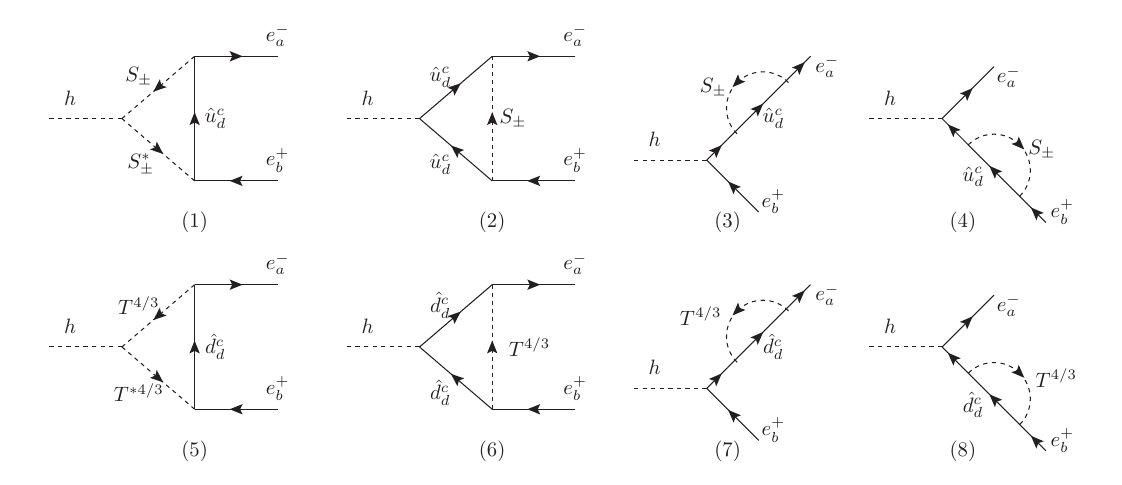}
	\caption{ One-loop Feynman diagrams with LQ exchanges contributing to LFV$h$ decay amplitudes predicted by the LQST framework, where $a, b, d=1,2,3$. }\label{f:hebea}
\end{figure}

The Yukawa  factors relating to couplings of the SM-like Higgs boson with up-type and down-type quarks $g^{L(R)}_{\hat{q}^c_i\hat{q}^c_i}$ derived  by identifying from the general part $\mathcal{L}^Y=-h\sum_{i=1}^3\hat{q}^C_{i} \left[ g^{L}_{\hat{q}^c_i\hat{q}^c_i} P_L + g^{R}_{\hat{q}^c_i\hat{q}^c_i} P_R\right]\hat{q}^C_{i} +\mathrm{h.c.}$, with $q=u, d$. In particular, these couplings are from mass Lagrangian in SM that mean is $g^{L}_{\hat{q}^c_i\hat{q}^c_i}=g^{R}_{\hat{q}^c_i\hat{q}^c_i}= gm_{q_i}/(2m_W)$. The coupling factor $\lambda_{hBB}$ was derived from the Higgs potential given in Appendix \ref{app:higgs} and shown in Eq.~\eqref{coup_hebea}, namely:
\begin{align}\label{coup_hebea}
\lambda_{hS_-S^*_-}=&-\frac{1}{2}v \left[s_{\theta }{}^2(2 \lambda _{13}^{\phi }+\lambda_{23}^{\phi }) +2c_{\theta }{}^2 \lambda _1^{\phi } +2\sqrt{2} \lambda  c_{\theta } s_{\theta } \right],
\crn \lambda_{hS_+S^*_+}=&-\frac{1}{2} v \left[c_{\theta }{}^2 (2 \lambda _{13}^{\phi }+\lambda _{23}^{\phi }) +2 s_{\theta }{}^2 \lambda_1^{\phi } -2 \sqrt{2} \lambda  c_{\theta } s_{\theta }\right],
\crn \lambda_{hS_-S^*_+}= & \lambda_{hS_+S^*_-}= -\frac{1}{2} v \left[c_{\theta } s_{\theta } (2 \lambda _{13}^{\phi }-2 \lambda _1^{\phi }+\lambda _{23}^{\phi })+\sqrt{2} \lambda  \left(c_{\theta}{}^2 - s_{\theta}{}^2\right)\right], 
\crn \lambda_{hTT}=& -v (\lambda _{13}^{\phi }+\lambda _{23}^{\phi }).
\end{align}
According to the results in Eq.~\eqref{coup_hebea}, we obtain the one-loop Feynman diagram corresponding to these contributions to LFV$h$ decays in this model is presented in Fig.~\ref{f:hebea}.

 The branching ratio (Br) of LFV$h$ decays are given by \cite{Pilaftsis:1992st, Arganda:2004bz, Arganda:2014dta} :
 \begin{equation}
 \mathrm{Br} (h \rightarrow e_be_a)\equiv \frac{\Gamma (h\rightarrow e_b^{+} e_a^{-})+\Gamma (h \rightarrow e_b^{-} e_a^{+})}{\Gamma^{\mathrm{total}}_{h}}
 \simeq   \fr{ m_{h}}{8\pi}\left(\vert \Delta^{(ab)}_L\vert^2+\vert \Delta^{(ab)}_R\vert^2\right), \label{eq_LFVwidth}
 \end{equation}
 where $\Gamma^{\mathrm{total}}_{h}\simeq 4.1\times 10^{-3}$ GeV \cite{LHCHiggsCrossSectionWorkingGroup:2016ypw}  and  $\Delta^{(ab)}_{L,R}$ are one-loop contribution arising from LQ exchange in Fig. \ref{f:hebea}.  The particular analytic formulas are derived from general results shown in Ref. \cite{Hue:2024rij}, namely 
\begin{align}
 \label{eq:deS}
 \Delta^{(ab)}_{L(R)}=&\Delta^{(ab)q\Phi\Phi'}_{L(R)}+ \sum_{\Phi}\Delta^{(ab)\Phi qq}_{L(R)},
\end{align}
where 
 \begin{align}
\label{eq:DeqPhiPhi}
\Delta^{(ab)q\Phi\Phi'}_L=& \sum_{i=1}^3\sum_{\Phi,\Phi'=S_\pm}  \frac{3 \lambda_{h\Phi\Phi'^*}}{16\pi^2}  \left[ m_{u_i}{g^{RL}_{ab,i\Phi \Phi'}} C_0   -\left(g^{LL}_{ab,i\Phi \Phi'} m_a C_1 +g^{RR}_{ab,i\Phi \Phi'}m_b C_2\right) \right]\left(m_{u_i}^2,m_\Phi^2,m_{\Phi'}^2\right)
\crn &+ \sum_{i=1}^3  \frac{3 \lambda_{hTT}}{16\pi^2}  \left(-g^{LL}_{ab,iTT} m_a C_1 \right)\left(m_{d_i}^2,m_{T}^2,m_{T}^2\right)
\crn &+ \frac{3g}{32 \pi^2m_W(m_a^2-m_b^2)} 
\crn &\times \sum_{i=1}^3\sum_{{\Phi=S_\pm, T^{ 4/3}}} \left[ g^{LR}_{ab,i\Phi} m_am_b m_{q_i} \left(B^{(1)}_0 -B^{(2)}_0\right) +g^{RL}_{ab,i\Phi} m_{q_i} \left(m_b^2 B^{(1)}_0 -m_a^2 B^{(2)}_0\right)
\right. \crn & \hspace{2.9cm}-\left. m_am_b \left(g^{LL}_{ab,i\Phi} m_b + g^{RR}_{ab,i\Phi} m_a \right)  \left(B^{(1)}_1 -B^{(2)}_1\right)\right] \left(m_{q_i}^2,m_\Phi^2\right),
\crn 	\Delta^{(ab)q\Phi\Phi'}_R= &	\Delta^{(ab)q\Phi\Phi'}_L\left[g^{LL} \leftrightarrow g^{RR},  g^{RL} \leftrightarrow g^{LR}\right],
\end{align}
 where $q=u,d$. The couplings $g^{XY}_{ab,i\Phi}$ are given in Eq. \eqref{eq:gXLR}, and $g^{XY}_{ab,i\Phi \Phi'}=  \Gamma_{ai\Phi}^{X*} \Gamma_{bi\Phi'}^{Y}$ with $X,Y=L,R$. More ever, all expressions in Eq. \eqref{eq:DeqPhiPhi} are written in terms of  well-known  Passarino-Veltman (PV) functions \cite{Passarino:1978jh}:  $C_{\delta}(m_{q_i}^2,m_\Phi^2,m_{\Phi'}^2)=C_{\delta}(m_a^2,m_h^2,m_b^2; m_{q_i}^2,m_\Phi^2,m_{\Phi'}^2)$ with $\delta=0,1,2$;  and $B^{(k)}_{0,1}(m_{q_i}^2,m_\Phi^2)=B^{(k)}_{0,1}(p_k^2;m^2_{q_i},m^2_\Phi)$ ($k=1,2$), using notations defined precisely in Ref. \cite{Hue:2024rij}, based on LoopTools \cite{Hahn:1998yk} implemented in our numerical investigation.    
 
As a result, $\Delta^{(ab)\Phi qq}_{L(R)}$ has the following simple formulas:
   \begin{align}
  \label{eq:DeSuu}
    \Delta^{(ab)\Phi qq}_L=&\sum_{i=1}^3\sum_{\Phi} \frac{3g  m_{q_i}}{ 32\pi^2 m_W} 
         \left[  g^{RL}_{ab,i\Phi} \left( B^{(12)}_0 +(m_{q_i}^2+ m_{\Phi}^2) C_0 +m_a^2 C_1 +m_b^2 C_2 \right) +g^{LR}_{ab,i\Phi} m_am_bX_0
  \frac{}{}\right. \crn&  \left. \frac{}{}\hspace{3.2cm}  +  m_{q_i} \Big((g^{LL}_{ab,i\Phi} m_a (C_0 +2C_1) +g^{RR}_{ab,i\Phi} m_b (C_0 +2C_2)\Big)  \right],
  \crn 	\Delta^{(ab)\Phi qq}_R= &	\Delta^{(ab)\Phi qq}_L\left[ g^{LL}_{ab,i\Phi} \leftrightarrow g^{RR}_{ab,i\Phi},  g^{RL}_{ab,i\Phi} \leftrightarrow g^{LR}_{ab,i\Phi} \right],
  \end{align}
  where $\Phi=S_\pm,T^{4/3}$;  and the PV-functions are  $C_{\delta}=C_{\delta}(m_a^2,m_h^2,m_b^2; m_\Phi^2,m_{q_i}^2, m_{q_i}^2)$ with $\delta=0,1,2$; and $B^{(12)}_{0}=B_{0}(m_h^2;m^2_{q_i},m^2_{q_i})$.  We can see that although both $\Delta^{q\Phi\Phi}_{L(R)}$ and $\Delta^{\Phi qq}_{L(R)}$ contain  divergences, the final sum of them is finite. In particular, the property of PV-functions \cite{Hue:2024rij}, and $g^{RR}_{abT^{4/3}}=g^{LR}_{abT^{4/3}}=g^{RL}_{abT^{4/3}}=0$ give:
  $$ \mathrm{div} \left[ \Delta^{q\Phi\Phi}_{L} \right] + \mathrm{div} \left[ {\Delta^{\Phi qq}_{L}} \right]\varpropto \mathrm{div}\left[B^{(12)}_0\right] +\frac{m_b^2 \mathrm{div}\left[B^{(1)}_0\right] -m_a^2 \mathrm{div}\left[B^{(2)}_0\right]}{m_a^2-m_b^2} =0. $$
 To complete the calculation of the LFV$h$ amplitudes and decay rates, we note that the terms proportional to $g^{LR(RL)}$ also contain factors of the quark masses. Consequently, these amplitudes can receive significant chirality-enhanced contributions from loops involving the top quark.

\subsection{\label{LFVZ} $Z\to e_b^\pm e_a^\mp$ decays}
\begin{figure}[ht]
	\centering 
	\includegraphics[trim=1cm 18.5cm 1cm 3cm, clip, width=1\textwidth]{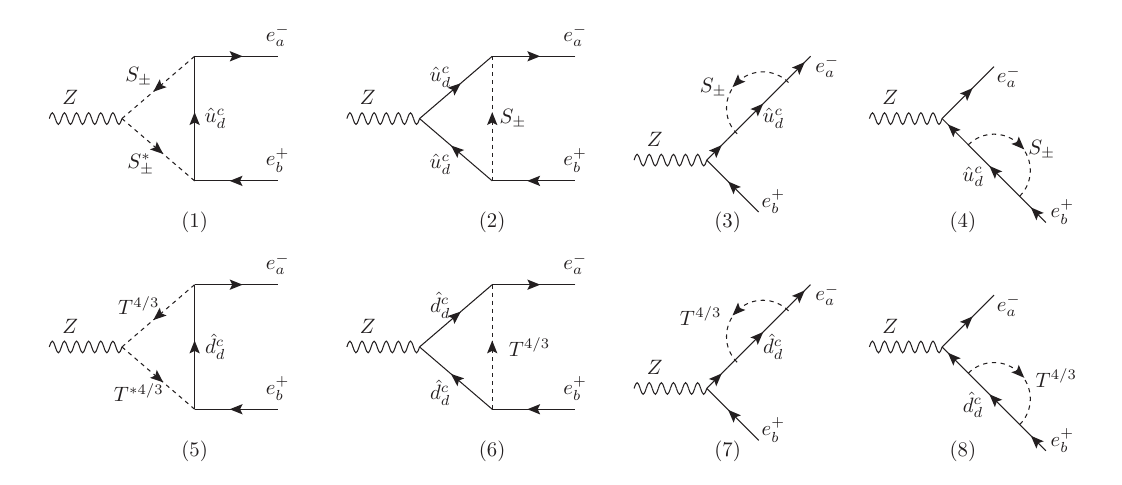}
	\caption{ One-loop Feynman diagrams with LQ exchanges contributing to LFV$Z$ decay amplitudes predicted by the LQST framework, where $a, b, d=1,2,3$ are the flavor indices of quarks. }\label{f:Zebea}
\end{figure}

The factor couplings $g_{Z\Phi\Phi}$ is derived from the kinetic terms
  $$\mathcal{L}_{\mathrm{kin}}^\Phi=(D_{\mu}\Phi)^*(D^{\mu}\Phi) =-ieg_{Z\Phi\Phi}Z_{\mu}\Phi\Phi^*(p_{\Phi^*}-p_\Phi)^{\mu}+\dots,$$
where $\Phi= S_\pm,\, T^{4/3}$. We get factor couplings $g_{Z\Phi\Phi}$, which is shown in Table.~\ref{tab_ZSS}
\begin{table}[ht]
	\centering 
	\renewcommand{\arraystretch}{1.5}
	\begin{tabular}{cccccc}
		\hline
		Vertex & factor &Vertex & factor &Vertex & factor \\
		\hline
		\hline
		$g_{ZS_-S^*_-} = g_{ZS_+S^*_+}$	& $\dfrac{t_W}{3} $&  $g_{ZTT}$
		& $\dfrac{3c^2_W+s^2_W}{3s_Wc_W} $ &	$g_{ZS_-S^*_+} = g_{ZS_+S^*_-}$
		& 0\\
		\hline			
	\end{tabular}
	\caption{The coupling factors of the $Z_\mu$ gauge boson and LQs.  \label{tab_ZSS} }\setlength{\tabcolsep}{10pt}
\end{table}
From results in Table.~\ref{tab_ZSS}, we consider the one-loop Feynman diagram corresponding to these contributions, which is presented in Fig.~\ref{f:Zebea}.

The Br of LFV$Z$ decays are $\mathrm{Br}(Z\to e^+_b e^-_a)= \Gamma (Z\to e^+_b e^-_a)/\Gamma_Z$, where the total decay width of the $Z$ boson is $\Gamma_Z= 2.4955$ GeV  \cite{ParticleDataGroup:2024cfk,ParticleDataGroup:2026mpi}, and \cite{Korner:1992an, DeRomeri:2016gum, Jurciukonis:2021izn, Hong:2023rhg}:
 \begin{align} 
 \label{eq_GAZeba}
 \Gamma (Z\to e^+_b e^-_a)= 	\frac{\sqrt{\lambda}}{16\pi m_Z^3}\times \left(\frac{e}{16\pi^2}\right)^2 \left( \frac{\lambda M_0}{12 m^2_Z} +M_1 +\frac{ M_2}{3 m^2_Z}\right),
 \end{align}
 where $\lambda= m^4_Z +m^4_{b} +m^4_{a} -2(m^2_Zm^2_{a} +m^2_Zm^2_{b} +m^2_{a}m^2_{b})$, 
 and the formulas of $M_{0,1,2}$ were given in Ref. \cite{Jurciukonis:2021izn}, which are presented in a reduced form \cite{Hong:2023rhg}
 \begin{align}
 \label{eq_Mi}
 M_0= & (m^2_Z -m_{a}^2 -m_{b}^2)\left(|\bar{b}^{\mathrm{LQST}}_L|^2 +|\bar{b}^{\mathrm{LQST}}_R|^2\right)  -4 m_{a} m_{b} \mathrm{Re}\left[ \bar{b}^{\mathrm{LQST}}_L  \bar{b}^{\mathrm{LQST}*}_R\right]
 \crn&
 - 4m_{b} \mathrm{Re}\left[ \bar{a}^{\mathrm{LQST}*}_R \bar{b}^{\mathrm{LQST}}_L   + \bar{a}^{\mathrm{LQST}*}_L \bar{b}^{\mathrm{LQST}}_R  \right] -  4m_{a}\mathrm{Re}\left[ \bar{a}^{\mathrm{LQST}*}_L \bar{b}^{\mathrm{LQST}}_L   + \bar{a}^{\mathrm{LQST}*}_R  \bar{b}^{\mathrm{LQST}}_R  \right] , 
 \crn M_1 = & 4 m_{a}m_{b} \mathrm{Re}\left[\bar{a}^{\mathrm{LQST}}_L\bar{a}^{\mathrm{LQST}*}_R \right],
 \crn  M_2 = &  \left[ 2 m^4_Z - m_Z^2\left( m_{a}^2 + m_{b}^2\right) - \left( m_{a}^2 - m_{b}^2\right)^2  \right] \left( |\bar{a}^{\mathrm{LQST}}_L|^2 +|\bar{a}^{\mathrm{LQST}}_R|^2\right).
 \end{align}
 where we omit the LFV index $(ab)$ in the right handed side for simplicity. 
 
 The contributions from diagrams with pure scalar exchanges were shown previously in Ref. \cite{Hue:2024rij}.  Particular formulas of the amplitudes are written as follows.  Final results for form factors of all diagrams in Fig. \ref{f:Zebea} are
 \begin{align}
 \label{eq:abLR}
 \overline{a}^{\mathrm{LQST}}_{L(R)}= \bar{a}^{q\Phi\Phi}_{L(R)}+\bar{a}^{\Phi qq}_{L(R)},\;  \overline{b}^{\mathrm{LQST}}_{L(R)}= \bar{b}^{q\Phi\Phi}_{L(R)}+\bar{b}^{\Phi qq}_{L(R)},
 \end{align}
  where sum of one-loop contributions from six diagrams (1), (3), (4), (5), (7), and (8) result in the following form factors  
  \begin{align}
 %Check 02 March 2024:Ok
 \label{eq_ab4LR}	
 %\left(g_{ZSS^*} +g_{ZT^{4/3}T^{-4/3}}\right)
 \bar{a}^{q\Phi\Phi}_{L}=&- 6 \sum_{\Phi} \sum_{i=1}^{3}g_{Z\Phi\Phi} g^{LL}_{ab,i\Phi} C_{00}
 \crn & - \frac{ 3t_{L}}{m_a^2 -m_b^2}\sum_{i=1}^3\sum_{\Phi}g_{Z\Phi\Phi}  \left[  m_{q_i}  \left( m_a g^{RL}_{ab,i\Phi}  + m_b g^{LR}_{ab,i\Phi}   \right) \left(B^{(1)}_0 -B^{(2)}_0\right) 
 \right. \crn& \left. \hspace{3cm} -m_am_b g^{RR}_{ab,i\Phi}\left( B^{(1)}_1-B^{(2)}_1\right)   - g^{LL}_{ab,i\Phi}   \left(m_a^2B^{(1)}_1 -m_b^2 B^{(2)}_1 \right) 
 \right],
 \crn  \bar{b}^{q\Phi\Phi}_{L} =& -2 \sum_{i=1}^3 g_{Z\Phi\Phi *} \left[\left(m_a  g^{LL}_{ab,i\Phi} X_1  +  m_b  g^{RR}_{ab,i\Phi} X_2\right)   -m_{q_i} g^{RL}_{ab,i\Phi}  X_0  \right] ,
 \crn  \bar{a}^{q\Phi\Phi}_{R} =& \bar{a}^{q\Phi\Phi}_{L} \left[ t_L\to t_R, g^{LL}_{ab,i\Phi} \leftrightarrow g^{RR}_{ab,i\Phi}, g^{RL}_{ab,i\Phi} \leftrightarrow g^{LR}_{ab,i\Phi} \right],
 \crn  \bar{b}^{q\Phi\Phi}_{R} =& \bar{b}^{q\Phi\Phi}_{L} \left[ g^{LL}_{ab,i\Phi} \leftrightarrow g^{RR}_{ab,i\Phi}, g^{RL}_{ab,i\Phi} \leftrightarrow g^{LR}_{ab,i\Phi} \right],
 \end{align}
 where  $\Phi=S_\pm, T^{4/3}$; $g^{XY}_{ab,i\Phi}$ is given in Eq. \eqref{eq:gXLR}, and  arguments for PV-funtions are $(m_{u_i}^2, m_{\Phi}^2, m_{\Phi}^2)$,  $B^{(k)}_{0,1}=B_{0,1}(p_k^2;m_{q_i}^2,m_\Phi^2)$.
 Forms factors corresponding to diagram (10) are
 \begin{align}
 \label{eq_ab6LR}	
 \bar{a}^{\Phi qq}_{L}=& -3 \sum_{\Phi}\sum_{i=1}^3 \left\{  g^{L}_{Z\hat{q}^c_i \hat{q}^c_i}\left[ g^{LL}_{ab,i\Phi}  m_{q_i}^2 C_0 + g^{RL}_{ab,i\Phi}  m_{a} m_{q_i}(C_0+C_1) \frac{}{}
 \right.\right.\crn&\left.\left.\qquad \qquad\qquad\qquad\quad
 +  g^{LR}_{ab,i\Phi}  m_{b} m_{q_i}(C_0+C_2) +  g^{RR}_{ab,i\Phi} m_{a} m_bX_0 \frac{}{}\right]   
 \right.\crn  &\left. \qquad\qquad\quad\; -g^{R}_{Z\hat{q}^c_i \hat{q}^c_i} \left[ g^{LL}_{ab,i\Phi} \left( (d-2)C_{00} +m_a^2 X_1 +m_b^2X_2 -m_Z^2 C_{12}\right) \frac{}{}
 \right.\right.\crn&\left.\left.\qquad \qquad\qquad\qquad\quad \frac{}{} +m_am_{q_i}g^{RL}_{ab,i\Phi}C_1 + m_bm_{q_i}g^{LR}_{ab,i\Phi}C_2 \right]\right\}
 \crn \bar{b}^{\Phi qq}_{L}=& -6 \sum_{\Phi}\sum_{i=1}^3 \left[\frac{}{} g^{L}_{Z\hat{q}^c_i \hat{q}^c_i}  \left( g^{RL}_{ab,i\Phi} m_{q_i}C_2 +  g^{RR}_{ab,i\Phi}  m_{b} X_2\right) 
 %
 %	\right.\crn&\left.   \qquad \; 
 +g^{R}_{Z\hat{q}^c_i \hat{q}^c_i}  \left(  g^{RL}_{ab,i\Phi}  m_{q_i} C_1 +  g^{LL}_{ab,i\Phi}  m_{a} X_1\right) 
 \right],
 \crn  \bar{a}^{\Phi qq}_{R}=& \bar{a}^{\Phi qq}_{L} \left[ g^{L}_{Z\hat{q}^c_i \hat{q}^c_i} \leftrightarrow g^{R}_{Z\hat{q}^c_i \hat{q}^c_i}, g^{LL}_{ab,i\Phi} \leftrightarrow g^{RR}_{ab,i\Phi}, g^{RL}_{ab,i\Phi} \leftrightarrow g^{LR}_{ab,i\Phi} \right],
  %%%%%%%%%%%%%%
 \crn\bar{b}^{\Phi qq}_{R} =& \bar{b}^{\Phi qq}_{L} \left[ g^{L}_{Z\hat{q}^c_i \hat{q}^c_i} \leftrightarrow g^{R}_{Z\hat{q}^c_i \hat{q}^c_i}, g^{LL}_{ab,i\Phi} \leftrightarrow g^{RR}_{ab,i\Phi}, g^{RL}_{ab,i\Phi} \leftrightarrow g^{LR}_{ab,i\Phi} \right], 
 \end{align}
 where $g^{XY}_{ab,i\Phi}$ is given in Eq. \eqref{eq:gXLR} and   arguments for PV-funtions are $(m_\Phi^2,m^2_{\hat{q}_i},m^2_{\hat{q}_i})$.  The coupling factors $g^{L,R}_{Z\hat{q}^c_i\hat{q}^c_i}$ derived from the general form $\mathcal{L}^{Zff}=e Z^{\mu}\sum_{f}\overline{f} \gamma_{\mu}\left[g^{L}_{Zff} P_L + g^{R}_{Zff} P_R\right]f +\mathrm{h.c.}$, with particular case of the LQST model, see for example the detailed formulas in Ref. \cite{Hung:2019jue}. We consider here the $g_{Zff}$ couplings is listed in Table \ref{t:Zff} for the quarks and charged leptons, namely:
 \begin{table}[ht]
 \renewcommand{\arraystretch}{1.2}
 	\begin{tabular}{|c|c|c|}
 		\hline
 	f& $g^L_{Zff}$ & 	$g^R_{Zff}$  \\
 		\hline
 		$e_a$&$\frac{2s_W^2-1}{2s_Wc_W}=t_L$& $t_W=t_R$\\
  \hline 
  	$\hat{u}_i=u,c,t$ &$\frac{-1}{s_Wc_W}\left(-\frac{1}{2} +\frac{2}{3}s_W^2\right)$ &$\frac{2 t_W}{3}$\\
  \hline 
  $\hat{d}_i=d,s,b$ &$\frac{-1}{s_Wc_W}\left(\frac{1}{2} -\frac{1}{3}s_W^2\right)$ &$\frac{-t_W}{3}$\\
  \hline
 	\end{tabular}
 	\caption{Coupling factors associated with Feynman rules for  $Z$ couplings  with two fermions. 
 		\label{t:Zff}}
 \end{table} 
We find that the LFV$Z$ amplitudes and decay rates likewise contain terms proportional to products of $g^{LR(RL)}$ and quark masses. Consequently, these amplitudes can receive significant chirality-enhanced contributions from loops involving the top quark. Combining this observation with the discussion of the cLFV and LFV$h$ decay rates, we qualitatively expect that chirality-enhanced contributions can dominate the LFV decay rates and $\Delta a_{e_a}$ when these observables become sufficiently large to approach their current experimental upper limits. The numerical analysis presented below examines these properties in greater detail.

\section{\label{sec_numerical} Numerical discussion}

Current searches for scalar LQs place the lower bounds on their masses at approximately \cite{CMS:2018svy,CMS:2018lab,CMS:2018ncu,ATLAS:2019ebv,ATLAS:2020dsk,ATLAS:2023prb}: $M_{T^{4/3}},\, M_{S_\pm} \geq 1.8~\mathrm{TeV}$. Throughout this work, we  investigate a general non-degenerate LQ mass spectrum with $2~\mathrm{TeV} \leq M_{T^{4/3}}, M_{S_\pm}\leq 8~\mathrm{TeV}$, which safely satisfies the current experimental lower limits.

The experimental values  used throughout our numerical analysis are taken from the latest Particle Data Group review \cite{ParticleDataGroup:2026mpi}, including the SM particle masses, the quark mixing matrix $V_{\mathrm{CKM}}$, and the gauge coupling constants. Neglecting the neutrino masses, we only show here the experimental values of charged fermion masses in the SM that are relevant to our continued numerical investigation below, which are taken from Ref.~\cite{ParticleDataGroup:2026mpi}, namely:
$m_e =5.1\times10^{-4}~\mathrm{GeV},\; m_\mu=0.105~\mathrm{GeV},\;
m_\tau=1.777~\mathrm{GeV},\;m_t=172.60~\mathrm{GeV}.$

The gauge interaction constants and related quantities are: $g =0.652,\; G_F=1.1664 \times 10^{-5} \mathrm{GeV},\; s_W^2=0.231,\; e^2 = 4\pi \alpha_{em} =\frac{4\pi}{137},\; m_W=80.3625~ \mathrm{GeV},\; m_Z=91.1879~ \mathrm{GeV}$. The scanning ranges of free parameters are chosen as follows:
\begin{align}
\label{eq:scanning}
	&|\lambda, \lambda^\phi_1, \lambda^\phi_{13},  \lambda^\phi_{23}| \leq 4\pi,\, |x^{L(R)}_{3i}|,|y^L_{3i}|\leq \sqrt{4\pi}\, \forall i=1,2,3,
\end{align}
where the quartic couplings satisfy the corresponding vacuum stability conditions and perturbativity requirements \cite{Bandyopadhyay:2016oif,Bandyopadhyay:2021kue,Chen:2022hle}. The Yukawa couplings are restricted to the perturbative regime and to the ranges compatible with the phenomenological constraints considered in Ref.~\cite{Bhaskar:2022vgk}.

\begin{figure}[ht!]
	%from dataLQ_ 13Sep26_R1.
	\centering
	\begin{tabular}{ccc}
		\includegraphics[width=5.7cm]{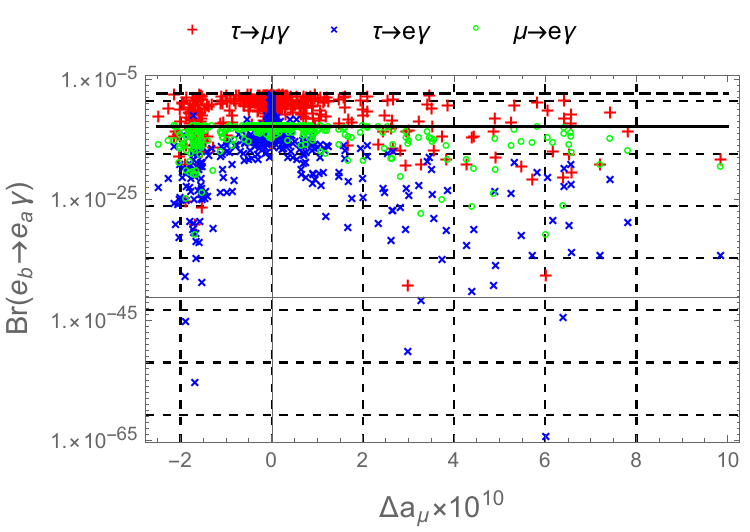} &
		\includegraphics[width=5.5cm]{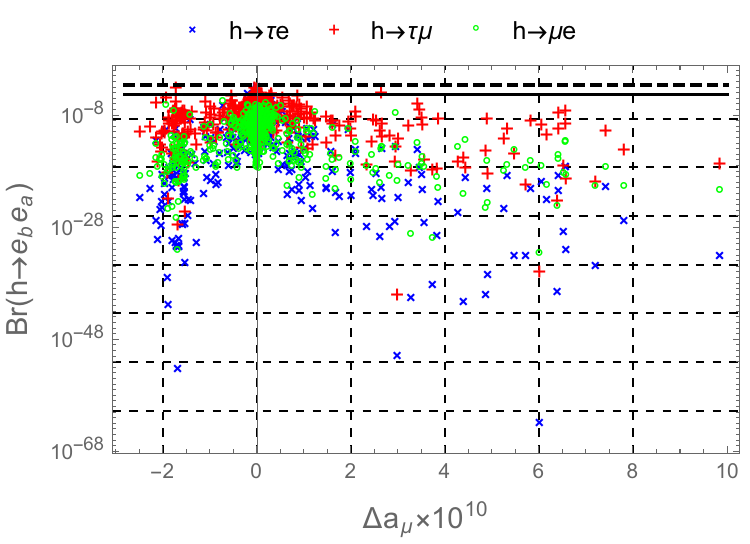} &
		\includegraphics[width=5.5cm]{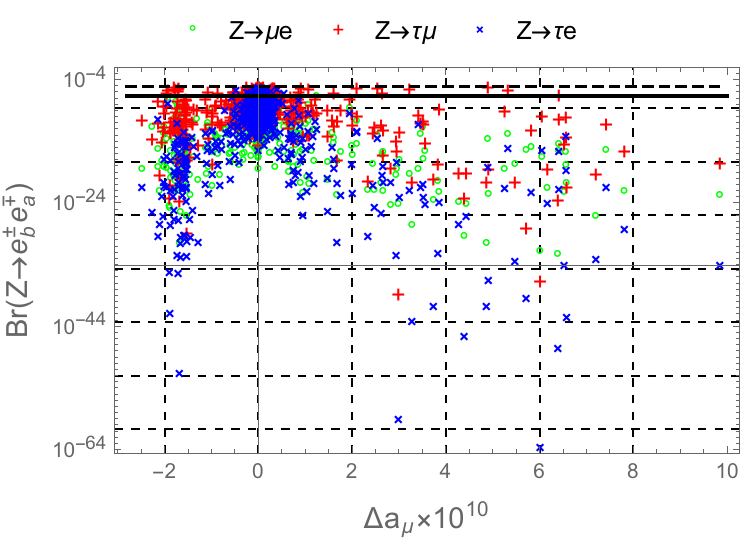} \\
		\includegraphics[width=5.5cm]{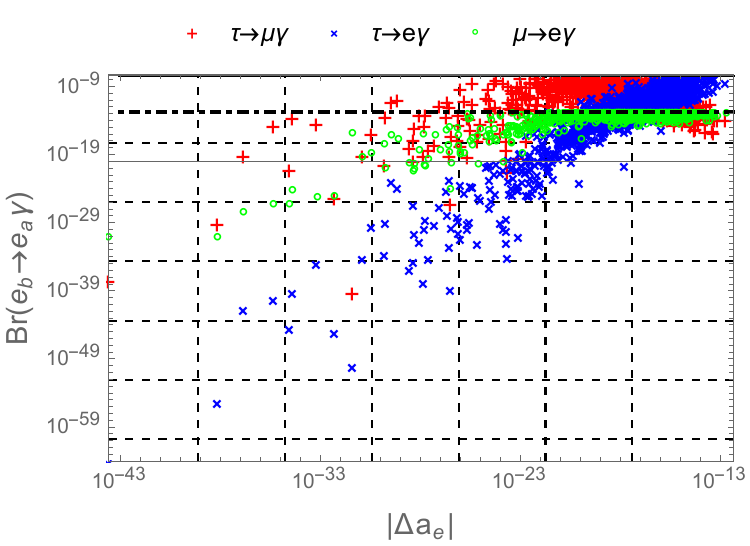} &
		\includegraphics[width=5.5cm]{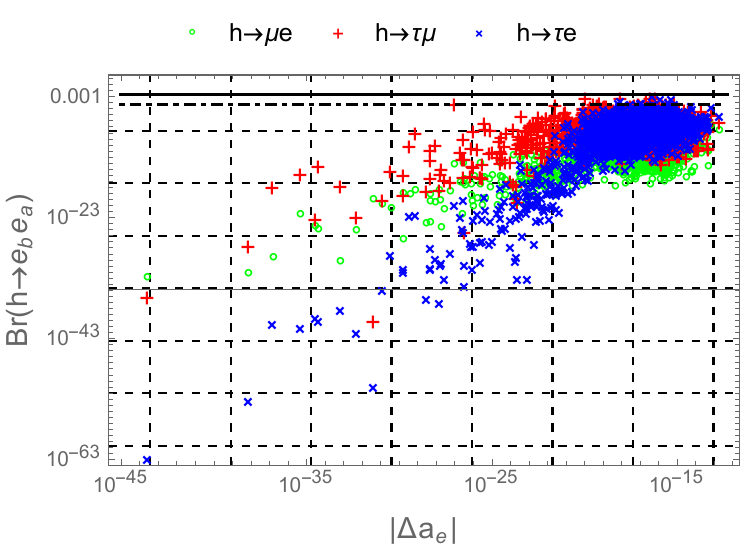} &
		\includegraphics[width=5.5cm]{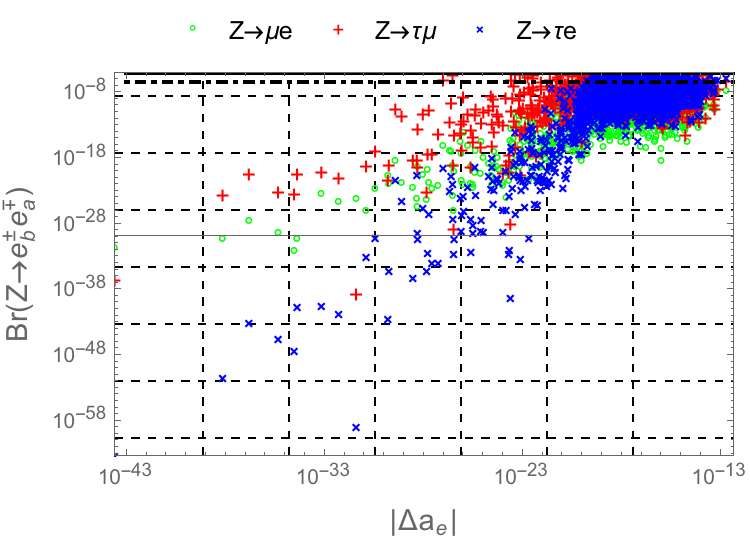} \\
	\end{tabular}
	\caption{LFV decay rates  as functions of  $\Delta a_{\mu}$(upper) and $|\Delta a_{e}|$ (lower) in case (1). In each panel, the two dashed and thick black horizontal lines denote the current experimental upper bounds in Eq.~\eqref{LFV_exp}. The corresponding $\tau\to\mu\gamma,\; h\to\tau e,\; Z\to\tau\mu$ for dashed line and $\mu\to e\gamma,\; h,Z\to\mu e$ for thick line in each panel, respectively.}\label{fig_LFVaxG}
\end{figure}

{
The numerical investigation focuses on the regions of parameter space that accommodate promising signals of LFV processes and $e_a$AMMs in forthcoming experiments. We impose three additional conditions on the scanned parameter points:

\begin{enumerate}
	\item We perform a general scan requiring sizable values of at least one of the AMM deviations,  $|\Delta a_e|>10^{-16}$ or $|\Delta a_\mu|>10^{-14}$, together with lower bounds of Br$(\mu \to e\gamma)>10^{-30}$. The lower bound excludes parameter points that yield no potentially observable LFV signals. 
	
\item We focus on regions with sizable values of $\Delta a_\mu$ and Br$(\mu \to e\gamma)$:  $ |\Delta a_{\mu}|>10^{-15}$ and Br$(\mu \to e \gamma) >10^{-15}$.
		
\item We focus on regions   with sizable values of $\Delta a_e$ and Br$(\mu \to e\gamma)$:  $5\times 10^{-13}\ge |\Delta a_{e}|>10^{-15}$ and Br$(\mu \to e \gamma) >10^{-15}$. 
\end{enumerate}
}
We first consider case (1), as shown in Fig.~\ref{fig_LFVaxG}, which illustrates the dependence of the LFV decay rates on $\Delta a_{e,\mu}$.
We find that all LFV decay rates can reach their current experimental sensitivities, but they cannot do so simultaneously, subject to the stringent constraint on the Yukawa coupling $|x^R_{31}|<0.1$ corresponding to $\Delta a_{e}\leq 2.1 \times 10^{-13}$. Interestingly, large values of $|\Delta a_{\mu}|= \mathcal{O}(10^{-10})$ result in small LFV decay rates for Br$(\tau \to e\gamma)$, and Br$(h,Z\to e_be_a)$. In contrast, large values of  $|\Delta a_{e}|$ still allow the decay rates to reach values close to their current experimental sensitivities.

These features arise from the stringent experimental constraint on Br$(\mu \to e\gamma)$  and the dominance of the chirality-enhanced contributions discussed in Eq.~\eqref{eq:cLFV1}. Namely, large $|\Delta a_{\mu}|$ drives Br$(\mu \to e\gamma)$ and Br$(\tau \to \mu \gamma)$  close to their experimental upper limits and consequently requires smaller values of $|\Delta a_{e,\tau}|$, leading to a suppressed Br$(\tau \to e\gamma)$.

\begin{figure}[ht!]
	\centering
	\begin{tabular}{ccc}
		\includegraphics[width=7.5cm]{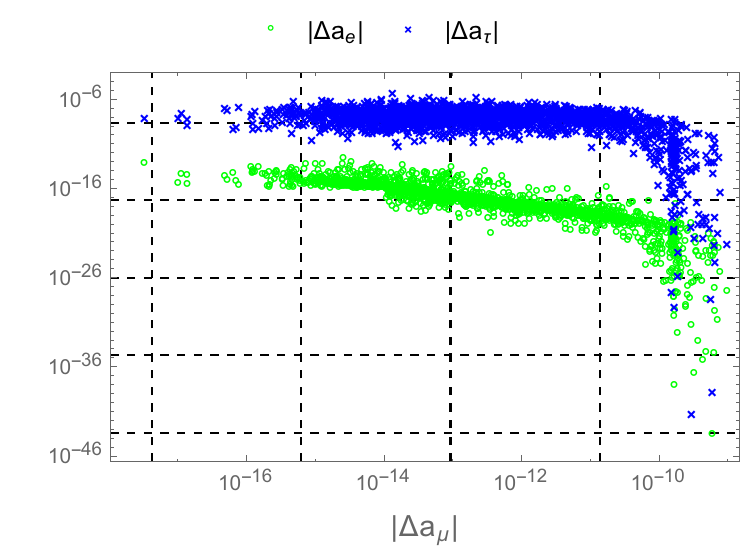} &
	\includegraphics[width=7.7cm]{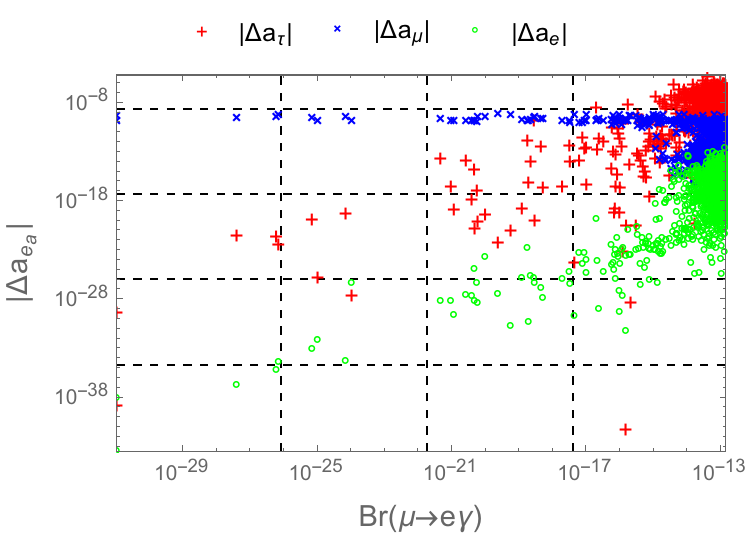} 	&
	 \\
		\end{tabular}
	\caption{$|\Delta a_{e_a}|$ as functions of $|\Delta a_{\mu}|$ and Br$(\mu \to e \gamma)$}\label{fig_axG}
\end{figure}

The correlations between  $|\Delta a_{\mu,\tau}| $  and $\Delta a_e$ are shown in Fig.~\ref{fig_axG},
%from dataLQ_ 13Sep26_R1.
confirming the qualitative behavior between $\Delta a_{\mu}$ and the LFV decay rates discussed above. In the left panel of Fig.~\ref{fig_axG}, $|\Delta a_{\mu,e}|$ cannot simultaneously attain the sizable values $|\Delta a_{e}|=\mathcal{O}(10^{-13})$ and $|\Delta a_{\mu}|=\mathcal{O}(10^{-10})$. Namely, for $|\Delta a_{\mu}|=\mathcal{O}(10^{-10})$, one obtains $|\Delta a_{e}|<\mathcal{O}(10^{-17})$, whereas $|\Delta a_{e}|=\mathcal{O}(10^{-13})$  implies $|\Delta a_{\mu}|<\mathcal{O}(10^{-14})$. The right panel shows that large values of $|\Delta a_{\mu}|$ are allowed for small Br$(\mu \to e\gamma)$, whereas $|\Delta a_{e}|\geq \mathcal{O}(10^{-14})$ is possible only for sufficiently large values of Br$(\mu \to e\gamma)$.

\begin{figure}[ht!]
	\centering
	\begin{tabular}{ccc}
		\includegraphics[width=5.5cm]{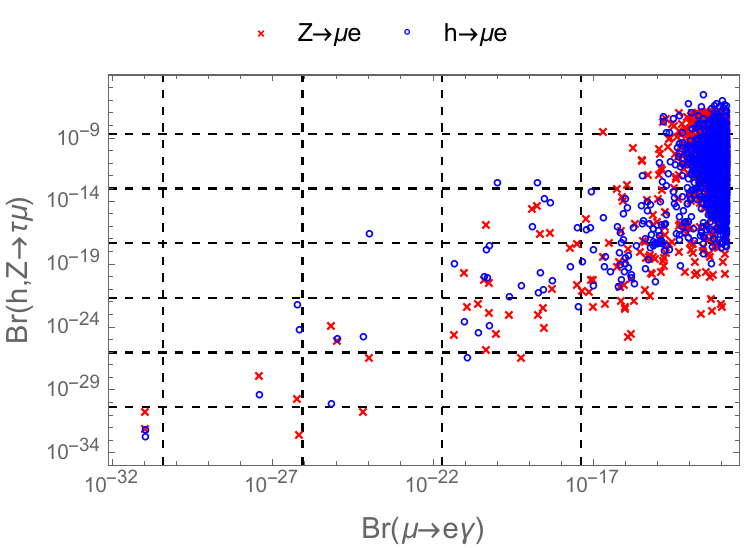}	&	\includegraphics[width=5.5cm]{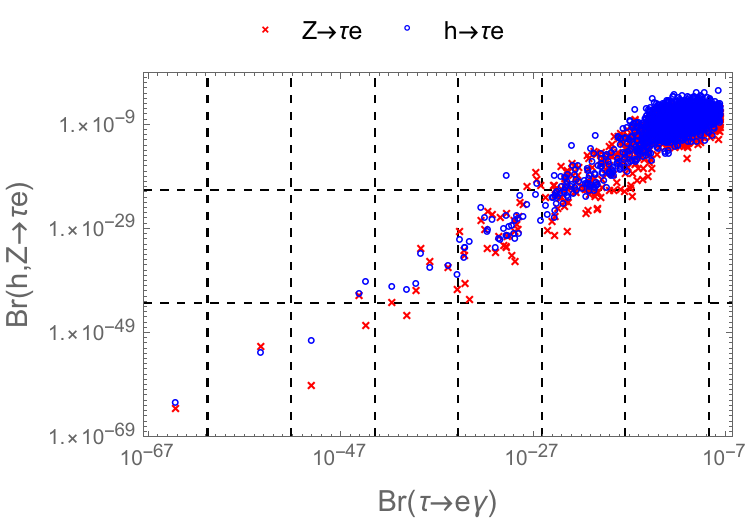} &
			\includegraphics[width=5.3cm]{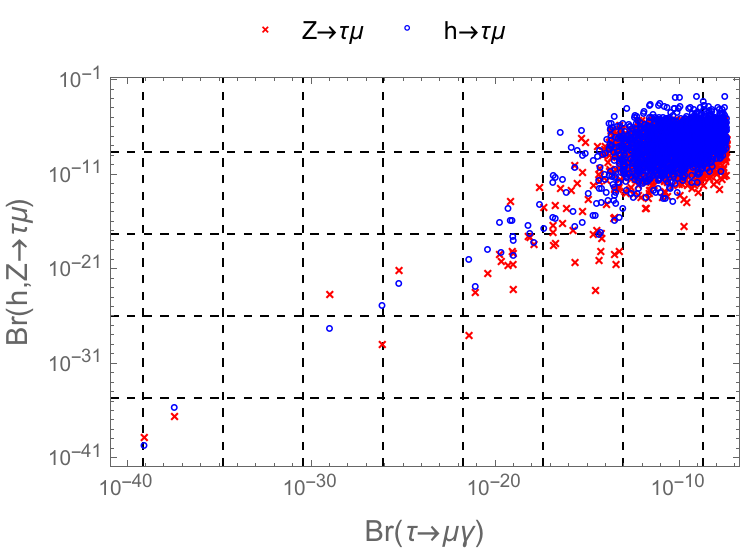} \\
	\end{tabular}
	\caption{The relationship between $\mathrm{Br}(h,Z\to e_be_a)$ with respect to $\mathrm{Br}(e_b\to e_a\gamma)$. }\label{fig_hZtemu}
\end{figure}

Next, Fig.~\ref{fig_hZtemu} shows the dependence of the Br$(h,Z\to e_b e_a)$  on $\mathrm{Br}(e_b\to e_a\gamma)$. 
It can be seen that the $h,Z\to e_b e_a$ decay channels exhibit correlations with their corresponding radiative decays $e_b\to e_a\gamma$, particularly when $\mathrm{Br}(e_b\to e_a\gamma)$ is sufficiently small.  In particular, the $h,Z\to\tau e$ channels show the strongest correlations, with their branching ratios exhibiting an approximately linear dependence on $\mathrm{Br}(\tau\to e\gamma)$. Therefore, the absence of a significant signal in the cLFV decay $e_b\to e_a\gamma$ would also imply that the $h,Z\to e_b e_a$ decay channels are unlikely to be observable.

\begin{figure}[ht!]
	\centering
	\begin{tabular}{cc}
%dataLQ_g2mu _ 12Sep26
		\includegraphics[width=7.5cm]{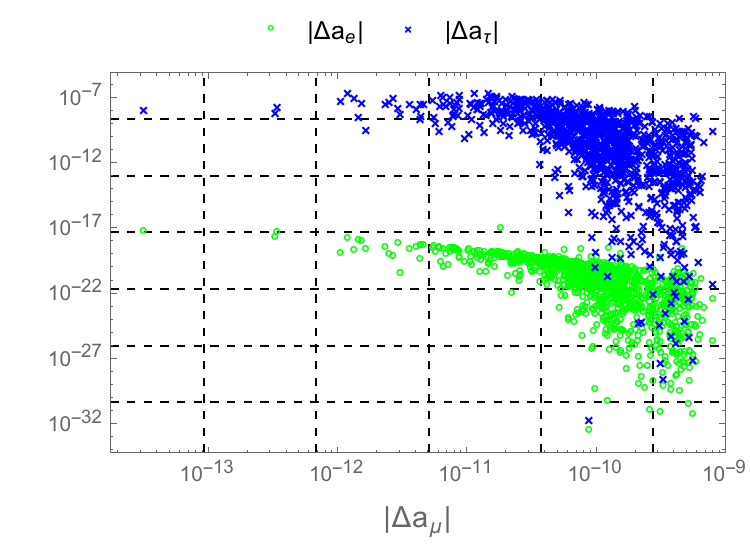} &
	\includegraphics[width=7.5cm]{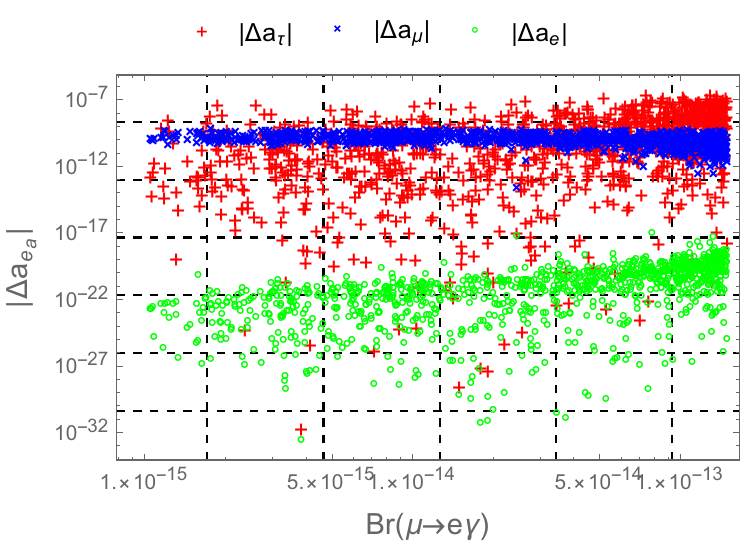}
 	\end{tabular}
	\caption{$|\Delta a_{e_a}|$ as functions of $|\Delta a_{\mu}|$  and Br$(\mu \to e \gamma)$ corresponding to  case (2).}\label{fig_axgm}
\end{figure}

We consider case (2), characterized by large  $|\Delta a_{\mu}|$. Figure~\ref{fig_axgm} shows the LFV decay rates as functions of  $\Delta a_{\mu}$ and $|\Delta a_e|$.
 In this case the maximal values of LFV decay rates and $\Delta a_{e_a}$ are as follows: $|\Delta a_{e}|\leq \mathcal{O}(10^{-18})$, $\Delta a_{\mu}\leq 8.5\times 10^{-10}$, $|\Delta a_{\tau}|\leq \mathcal{O}(10^{-7})$, Br$(\mu \to e \gamma) \leq 1.5\times 10^{-13}$,  Br$(\tau \to e \gamma) \leq  \mathcal{O}(10^{-12})$, Br$(\tau  \to \mu \gamma) \leq 4.2\times 10^{-8}$, Br$(h  \to  \mu e) \leq 1.9\times 10^{-5}$, Br$(h  \to \tau e ) \leq  \mathcal{O}(10^{-6})$, Br$(h  \to \tau \mu ) \leq 1.6\times 10^{-3}$, Br$(Z  \to  \mu e) \leq 8.5\times 10^{-8}$, Br$(Z  \to \tau e ) \leq 4.0\times 10^{-6}$, and Br$(Z  \to \tau \mu ) \leq 6.5\times 10^{-6}$. The  relevant Yukawa couplings are also stringent: $|y^{L}_{31}|, |x^{L}_{31}|<0.1$ and $|x^{R}_{31}|<0.01$. Therefore, when $|\Delta a_{\mu}|>10^{-15}$,  Br$(\tau \to e \gamma)$ and Br$(h  \to \tau e ,\mu e)$ are much smaller than the forthcoming experimental sensitivities. In contrast, the remaining LFV decay channels could still provide promising signals for future experimental searches.

\begin{figure}[ht!]
	\centering
	\begin{tabular}{cc}
	%dataLQ_gemax _ 12Sep26R1
		\includegraphics[width=7.5cm]{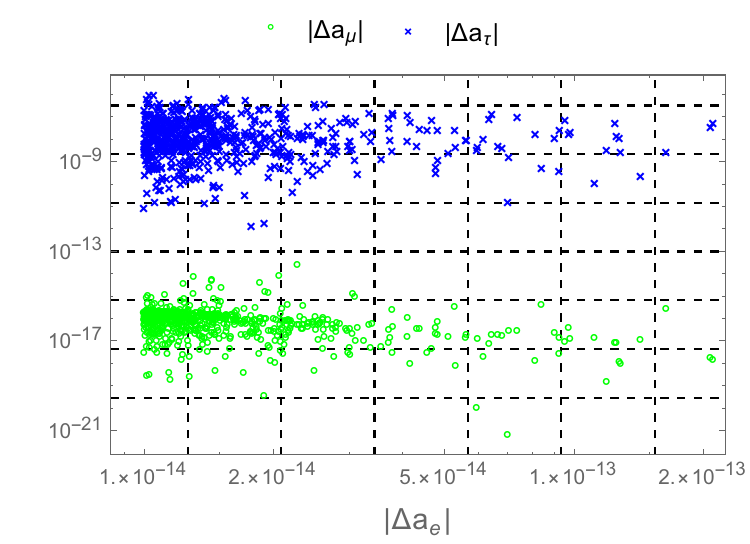} 
		&
		\includegraphics[width=7.5cm]{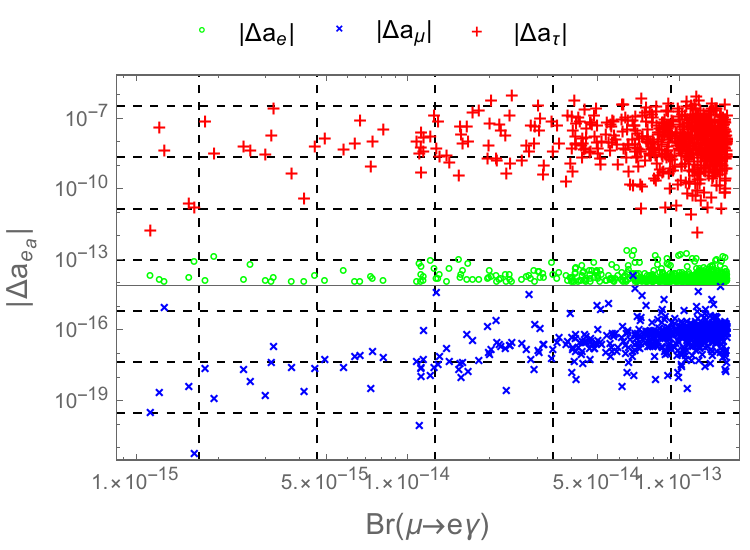} 
		\\
	\end{tabular}
	\caption{$|\Delta a_{e_a}|$ as functions of $|\Delta a_{e}|$ and Br$(\mu \to e \gamma)$ corresponding to case (3).}\label{fig_ax}
\end{figure}

We next consider case (3), characterized by large $|\Delta a_{e}|$. The corresponding LFV decay rates are shown in Fig.~\ref{fig_ax}. 
The maximal values of the LFV decay rates and AMM deviations are as follows: $|\Delta a_{e}|\leq \mathcal{O}(10^{-13})$, $|\Delta a_{\mu}|\leq \mathcal{O}(10^{-14})$, $|\Delta a_{\tau}|\leq 10^{-6}$, Br$(\mu \to e \gamma) \leq 1.5\times 10^{-13}$,  Br$(\tau \to e \gamma) \leq  3.3 \times 10^{-8}$, Br$(\tau  \to \mu \gamma) \leq \mathcal{O}(10^{-11})$, Br$(h  \to  \mu e) \leq \mathcal{O}(10^{-9})$, Br$(h  \to \tau e ) \leq  1.9 \times 10^{-3}$, Br$(h  \to \tau \mu ) \leq \mathcal{O}(10^{-7})$, Br$(Z  \to  \mu e) \leq 1.32\times 10^{-7}$, Br$(Z  \to \tau e ) \leq 5.0\times 10^{-6}$, and Br$(Z  \to \tau \mu ) \leq 2.4\times 10^{-7}$.  The  relevant Yukawa couplings are also stringent: $|y^{L}_{32}|, |x^{L}_{32}|<0.1$ and $|x^{R}_{32}|<0.005$. Now the suppressed LFV decay rates are Br$(\tau  \to \mu \gamma)$ and Br$(h \to \tau \mu, \mu e) $. 

\begin{figure}[ht!]
	\centering
	\begin{tabular}{cc}
		%dataLQ_g2mu _ 12Sep26
	\includegraphics[width=7.5cm]{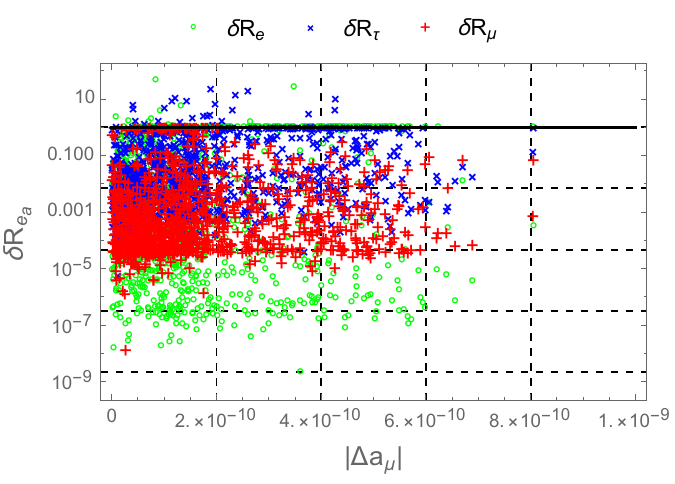} 
	&
	\includegraphics[width=7.5cm]{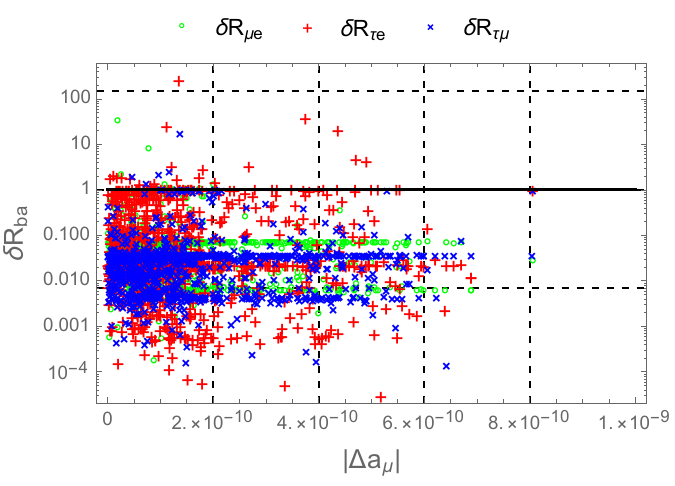} 
	\\
		%dataLQ_gemax _ 12Sep26R1
		\includegraphics[width=7.5cm]{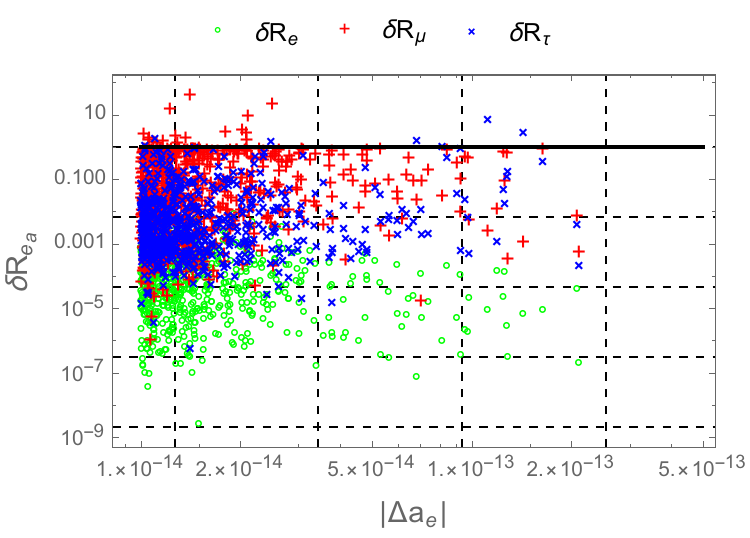} 
		&
		\includegraphics[width=7.5cm]{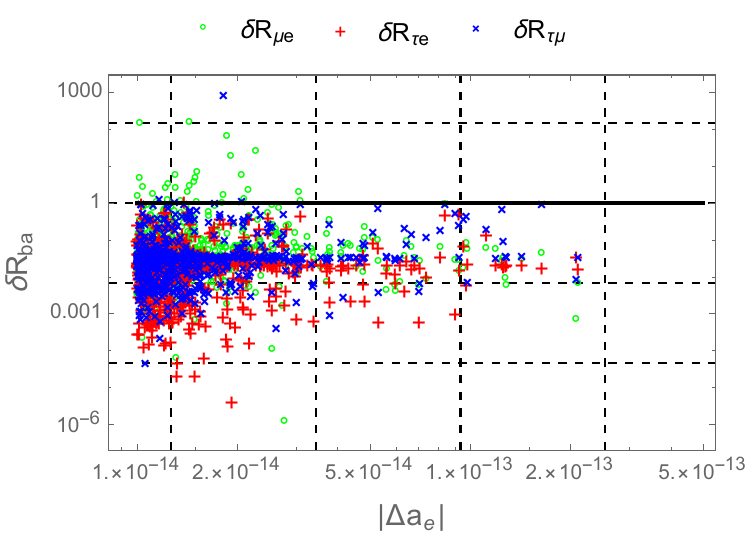} 
		\\
	\end{tabular}
	\caption{$\delta R_{e_a}$ and $\delta R_{ba}$ as functions of $|\Delta a_{\mu}|$ and $|\Delta a_{e}|$  corresponding to cases (2) and (3) respectively .}\label{fig:ae_Rba}
\end{figure}

Finally, we investigate the conditions under which the approximate formulas given in Eqs.~\eqref{eq:aeaLR}, \eqref{eq:emuLFV}, and \eqref{eq:cLFVLR} are valid, with the aim of providing simple estimates when new experimental data become available. We define the relative deviations between the numerical results obtained from the approximate and full expressions as
$$\delta R_{e_a}= \left| 1- \left| \frac{\Delta a^{\mathrm{LR}}_{e_a}}{\Delta a_{e_a}}\right|\right| ,\; \delta R_{ba} = \left| 1-  \frac{\mathrm{Br}(e_b\to e_a \gamma)^ \mathrm{LR}}{\mathrm{Br}(e_b\to e_a \gamma)}\right|.$$
Here, $\delta R_{e_a}=0$ and $\delta R_{ba}=0$ indicate that the chirality-enhanced contributions dominate the corresponding observables. Figure~\ref{fig:ae_Rba} shows the dependence of these deviations on the large values of $|\Delta a_\mu|$ and $|\Delta a_e|$, corresponding to cases (2) and (3), respectively.
First, $\delta R_{\tau}$ depends only weakly on both $|\Delta a_{\mu}|$ and $|\Delta a_{e}|$, consistent with the fact that the LR coupling product relevant to this quantity is independent of the requirements of large $\Delta a_{e,\mu}$ and Br$(\mu \to e\gamma)$.  In contrast, large values of $|\Delta a_{\mu}|\ge 7\times 10^{-10}$  require small values of $\delta R_{\mu}<0.01$ and $\delta R_{\mu e}<0.05$. For smaller values of  $|\Delta a_{\mu}|<3\times 10^{-10}$, larger values of $\delta R_{\mu}$ and $\delta R_{\mu e}$, reaching $\mathcal{O}(1)$, are allowed, implying that the chirality-enhanced contributions may no longer dominate. A similar behavior is observed for large values of $|\Delta a_e|\geq10^{-13}$, for which $\delta R_e<10^{-3}$ and $\delta R_{\mu e}<0.1$. Therefore, the approximate formulas given in Eqs.~\eqref{eq:aeaLR} and \eqref{eq:emuLFV} are valid for sufficiently large values of$|\Delta a_{e_a}|$. For small values of $|\Delta a_{e_a}|$, the interplay among different one-loop contributions, including the chirality-enhanced contributions, must be studied in detail.

\section{\label{conclusion} Conclusion}
In the LQST framework, we have performed a detailed study of the $e_a$AMMs, cLFV, LFV$h$, and LFV$Z$ decays. We have calculated the one-loop contributions and also performed a numerical analysis of the relevant parameter space and presented the results through several illustrative correlations among the considered observables. Our results show that the two LQs can provide viable contributions to the $e_a$AMMs while satisfying the current experimental constraints, particularly the stringent bound from $\mu\to e\gamma$. An important feature of our analysis is the correlation between the branching ratios of the $h,Z\to e_b e_a$ decays and $\mathrm{Br}(e_b\to e_a\gamma)$. In particular, the $h,Z\to\tau e$ channels exhibit an approximately linear correlation with $\mathrm{Br}(\tau\to e\gamma)$, indicating that LFV$h$ and LFV$Z$ decays can provide complementary probes of the parameter space associated with the corresponding radiative decays. Furthermore, the model allows several LFV channels, including $\mu\to e\gamma$, $\tau\to\mu\gamma$, $h\to \tau e$, and $Z\to\mu^\pm e^\mp, \tau^\pm\mu^\mp$, to approach their current experimental sensitivities. These correlations among the $e_a$AMMs and LFV observables, together with the possibility of sizable LFV signals, provide complementary probes of the LQST parameter space in future searches.

\section*{Acknowledgments}
We are grateful to Prof. Maxim KHLOPOV for his helpful comments. N.H.T. Nha was funded by the Master, PhD Scholarship Programme of Vingroup Innovation Foundation (VINIF), code VINIF.2025.TS24.

\appendix

\section{\label{app:higgs} Higgs bosons}
In this work, the Higgs potential respects the generalized lepton number introduced in Ref.~\cite{Dorsner:2019itg}. This Higgs potential is more general than that considered in Ref.~\cite{Bhaskar:2022vgk}, namely:
\begin{align}
\label{eq_Vh}
V_h = & \mu_\phi^2(\phi^\dagger \phi) + \mu^2_{1}S^{*}S  + \mu^2_{3}\mathrm{Tr}(T^{\dagger}T)   + \lambda_\phi (\phi^\dagger \phi)^2  + \lambda_{1} \left(S^{*}S\right)^2  + \lambda_{1}^\phi \left(\phi^\dagger \phi\right) \left(S^{*}S\right)
\crn &+ \lambda_{13} \left[ \mathrm{Tr}(T^{\dagger}T)\right]^2  + \lambda_{23} \mathrm{Tr}\left[ (T^{\dagger}T)^2\right]  + \lambda_{33} \mathrm{Tr}\left(TT\right)   \mathrm{Tr}\left(T^*T^*\right) + \lambda_{13}^\phi  (\phi^\dagger \phi) \mathrm{Tr} (T^{\dagger}T) 
\crn & +  \lambda_{23}^\phi  \mathrm{Tr}\left[ \phi^\dagger T^{\dagger}T \phi   \right]
+ \left[ \lambda \mathrm{Tr} \left( \phi^\dagger T\phi\right) S^*+\mathrm{h.c.}  \right]
\end{align}
%
%\section{\label{app:higgs} Masses and mixing matrices of LQs}
The LQ mass matrix is derived from the Higgs potential $V_h$ in Eq.~\eqref{eq_Vh}. By expanding the potential and collecting the quadratic terms in the LQ fields, we obtain the corresponding mass terms:
\begin{align}
\label{eq:Lmass}
-\mathcal{L}^S_{\mathrm{mass}} =& M^2_{T} T^{\frac{4}{3}}T^{-\frac{4}{3}} + M^2_{T^{\frac{2}{3}}} T^{\frac{2}{3}}T^{-\frac{2}{3}}  + \left(S^{1/3}, T^{1/3} \right) \mathcal{M}^2_{ST}\begin{pmatrix}
	 S^{-1/3} &
 T^{-1/3}
\end{pmatrix}^T,
\end{align}
where  the fields $T^{\pm\frac{4}{3}}$ and $T^{\pm\frac{2}{3}}$ are directly identified as the physical states with masses 
\begin{align}
\label{eq:m2pm}
M^2_T=M^2_{T^{\frac{4}{3}}}=  & \mu _3^2 +  \frac{1}{2} v^2 (\lambda _{13}^{\phi }+\lambda _{23}^{\phi }), 
\; 	M^2_{T^{\frac{2}{3}}}=   \mu _3^2 + \frac{\lambda _{13}^{\phi } v^2}{2}. 
\end{align}
On the other hand,  $\mathcal{M}^2_{ST}$ is given by
 \begin{equation}
 \label{eq:M2S13}
 \mathcal{M}^2_{ST}= 
\left(
\begin{array}{cc}
	\mu _1^2 +\frac{\lambda _1^{\phi } v^2}{2} & \frac{\lambda  v^2}{2 \sqrt{2}} \\
	\frac{\lambda  v^2}{2 \sqrt{2}} & \mu _3^2 +\frac{1}{4} (2 \lambda _{13}^{\phi }+\lambda _{23}^{\phi }) v^2 \\
\end{array}
\right). 
 \end{equation}

The mass-squared matrix $\mathcal{M}^2_{ST}$ is diagonalized by the matrix $U_{ST}$ as
 \begin{equation}
 \label{eq:US}
 U_{ST}=\begin{pmatrix}
 	c_{\theta}& s_{\theta} \\
 	-s_{\theta}& c_{\theta}
 \end{pmatrix} \to U_{ST} \mathcal{M}^2_{ST}U_{ST}^T= \mathrm{diag}\left( M^2_{S_-},\; M^2_{S_+}\right),
 \end{equation}
where $s_\theta = \text{sin}\theta$, $c_\theta = \text{cos}\theta$ satisfying
\begin{align}
\label{eq:t2t}
t_{2\theta} \equiv \text{tan}(2\theta) =& \frac{2 \sqrt{2} \lambda  v^2}{v^2 (-2 \lambda _{13}^{\phi }+2 \lambda _1^{\phi }-\lambda _{23}^{\phi })+4 (\mu _1^2-\mu _3^2)}
\end{align}
with $\theta \in \left[ -\frac{\pi}{4}, \frac{\pi}{4}\right]$ according to Ref. \cite{Bhaskar:2022vgk}.  $M^2_{S_\pm}$ denote the squared masses of the physical states $S_{\pm}$: 
\begin{align}
 	\label{eq:M2Spm}
 	M^2_{S_-} =& \frac{1}{2c_{2\theta}}\left[ c_{\theta }^2 \left(\lambda _1^{\phi } v^2+2 \mu _1^2\right)  -s_{\theta }^2 \left(\lambda _{13}^{\phi } v^2+2 \mu _3^2\right) \right],
\crn  M^2_{S_+} = & \frac{1}{2c_{2\theta}}\left[ -s_{\theta }^2 \left(\lambda _1^{\phi } v^2+2 \mu _1^2\right)  +c_{\theta }^2 \left(\lambda _{13}^{\phi } v^2+2 \mu _3^2\right) \right].
 \end{align}
The relation between the initial  and the physical bases is given by
\begin{align}
\label{eq:STphy}
 S^{1/3} =c_{\theta} S_- - s_{\theta} S_+, \quad T^{1/3}=s_{\theta} S_- + c_{\theta} S_+.
\end{align}

%----------------

%%%%%%%%%%%%%%%%%%%%%%%%%%%%%%%%%%%%%%%%%%%%
\end{document}